\documentclass[12pt]{article}
\usepackage{epsf}

\def\lesssim{\mathrel{\mathpalette\vereq<}}
\def\gtrsim{\mathrel{\mathpalette\vereq>}}
\makeatletter
\def\vereq#1#2{\lower3pt\vbox{\baselineskip1.5pt \lineskip1.5pt
\ialign{$\m@th#1\hfill##\hfil$\crcr#2\crcr\sim\crcr}}}
\makeatother

\renewcommand{\theequation}{\thesection.\arabic{equation}}

\begin{document}
\begin{titlepage}
\begin{center}
\today     \hfill UCB-PTH-97/46\\
~{} \hfill LBNL-41028\\
~{} \hfill hep-ph/9711264\\

\vskip .1in

{\large \bf Next-to-Minimal Supersymmetric Standard Model\\
  with the Gauge Mediation of Supersymmetry Breaking}\footnote{This
  work was supported in part by the Director, Office of Energy
  Research, Office of High Energy and Nuclear Physics, Division of
  High Energy Physics of the U.S.  Department of Energy under Contract
  DE-AC03-76SF00098 and in part by the National Science Foundation
  under grant PHY-90-21139.  AdG was also supported by CNPq (Brazil).
  HM was also supported by Alfred P. Sloan Foundation.}

\vskip 0.3in

Andr\'e de Gouv\^ea, Alexander Friedland, and Hitoshi Murayama

\vskip 0.05in

{\em Theoretical Physics Group\\
     Ernest Orlando Lawrence Berkeley National Laboratory\\
     University of California, Berkeley, California 94720}

\vskip 0.05in

and

\vskip 0.05in

{\em Department of Physics\\
     University of California, Berkeley, California 94720}

\end{center}

\vskip .1in
\begin{abstract}
  We study the Next-to-Minimal Supersymmetric Standard Model (NMSSM)
  as the simplest candidate solution to the $\mu$-problem in the
  context of the gauge mediation of supersymmetry breaking (GMSB).  We
  first review various proposals to solve the $\mu$-problem in models
  with the GMSB.  We find none of them entirely satisfactory and point
  out that many of the scenarios still lack quantitative studies, and
  motivate the NMSSM as the simplest possible solution.  We then study
  the situation in the Minimal Supersymmetric Standard Model (MSSM)
  with the GMSB and find that an order $10$\% cancellation is necessary
  between the $\mu$-parameter and the soft SUSY-breaking parameters to
  correctly reproduce 
  $M_Z$.  Unfortunately, the NMSSM does not to give a
  phenomenologically viable
  solution to the $\mu$-problem.  We present quantitative arguments
  which apply both for the low-energy and high-energy GMSB and prove
  that the NMSSM does not work for either case.  Possible
  modifications to the NMSSM are then discussed.  The NMSSM with
  additional vector-like quarks works phenomenologically, but requires
  an order a few percent cancellation among parameters. We point out that this
  cancellation has the same origin as the cancellation required in the
  MSSM.
\end{abstract}

\end{titlepage}

\newpage

\section{Introduction}

The primary motivation for supersymmetry (SUSY) is to stabilize the
smallness of the electroweak scale against radiative corrections
\cite{SUSY1,SUSY2,SUSY3},
which can be as large as the Planck scale if the Higgs bosons are
truly elementary.  Once the electroweak scale is set in the
tree-level Lagrangian, it only receives logarithmic radiative
corrections, and hence its order of magnitude is not changed.  
Moreover, the electroweak symmetry remains
unbroken in the Minimal Supersymmetric Standard Model (MSSM) in the absence
of explicit SUSY-breaking
parameters.  Therefore, one can view the electroweak symmetry breaking
as being triggered by the soft SUSY breaking.  Indeed, the soft
SUSY-breaking mass-squared 
of the Higgs boson can be driven negative due to the top quark loop
\cite{radbreak} while
all the other scalar bosons still have positive mass-squared.  In this sense,
there is nothing special about the Higgs boson.  It is just
one of many scalar bosons, which happens to acquire a negative
mass-squared due to the top quark loop.  This idea eliminates one of
the 
least appealing features of the Standard Model.  However, there are
at least two open questions.  First, SUSY by itself does not explain
why the electroweak scale is small to begin with.  Therefore, SUSY
makes the smallness of the electroweak scale ``technically natural,''
but not truly natural.  Second, the MSSM contains one dimensionful
parameter (the $\mu$-parameter), allowed by SUSY, in the
superpotetial. The natural values of $\mu$ are either the Planck mass
(the only natural dimensionful parameter available) or zero, but
recent experimental constraints imposed by LEP2 imply that a nonzero
$\mu\lesssim 50$~GeV is required \cite{LEP2}.
   
SUSY, fortunately, can potentially explain the smallness of the
electroweak scale if it is broken dynamically \cite{SUSY2}.  The
perturbative non-renormalization theorem forbids the generation of a mass
scale in the superpotential if it is absent at the tree-level.
However, non-perturbative effects can violate the non-renormalization
theorem, and a mass scale can be generated by a dimensional
transmutation: $\Lambda_{\rm \begin{picture}(25,0)(0,0)
        \put(0,0){\scriptsize SUSY}
        \put(0,0){\line(4,1){22}}
        \end{picture}}\sim M_{\rm
Planck}e^{-8\pi^{2}/g^{2}|b_{0}|}$, if an
asymptotically free gauge theory is responsible for SUSY breaking.
There has been major progress in building models of dynamical SUSY
breaking \cite{DN, DNS, DNNS, dynSUSY, direct, Amess, direct2}, 
which became possible with
the detailed understanding of the
non-perturbative dynamics of SUSY gauge theories \cite{IS}.
Furthermore, the so-called gauge mediation of SUSY breaking (GMSB)
\cite{SUSY3,oldGM}  can
generate soft SUSY-breaking parameters in the SUSY Standard Model in a
phenomenologically desired form.  Therefore, there is hope of
understanding the smallness of the electroweak scale in a truly natural
manner.

However, the other question remains largely unanswered: how can the
dimensionful parameters in the superpotential naturally be of the order
of the SUSY-breaking parameters?  There have been extensive
discussions on this subject in the literature which we briefly
summarize in Section 2.  Unfortunately, many of the proposed mechanisms
rely on either small parameters, accidental cancellations, or the absence
of interactions allowed by symmetries.  We find the current
situation to be rather unsatisfactory.

A natural direction to follow is to start with a superpotential which
does not contain a dimensionful parameter and hope that the
electroweak scale is generated solely due to the soft SUSY-breaking
parameters.  The simplest model which can potentially work along this
line is the Next-to-Minimal Supersymmetric Standard Model (NMSSM) \cite{Eetal},
which replaces the $\mu$-parameter by the vacuum expectation value of
an electroweak singlet superfield.  We revisit this possibility with
detailed quantitative studies in this paper.  Unfortunately, our
conclusion is negative.  The NMSSM by itself does not produce a
phenomenologically viable electroweak symmetry breaking even if we
vary the messenger scale.  The major experimental constraints include
Higgs boson and slepton searches.  Certain simple modifications
can evade phenomenological constraints, but require a cancellation
among parameters accurate to a few percent.  We
present all of these points quantitatively in this paper, and hope
that our results prompt further investigations in understanding the
origin of the $\mu$-parameter in models with the GMSB.

The paper is organized as follows.  In the next section, we review the
situation of the $\mu$-problem in models with the GMSB, and discuss various
proposals to explain the origin of the $\mu$-parameter.  We, however,
find none of them entirely satisfactory.  Even if we accept one of the
proposed models, it is still necessary to check whether the generated
$\mu$-parameter is phenomenologically allowed.  We address this
question in Section~3, and find that the currently available
experimental lower bounds on superparticle masses already require a
cancellation of order $10$\% between the $\mu$-parameter and
soft SUSY-breaking parameters to reproduce the observed $M_{Z}$.  Then,
as the major 
part of our study, we present the quantitative results of the
electroweak symmetry breaking in the NMSSM with the GMSB in
Section~4 and find that there is no phenomenologically viable
parameter set even if we vary the messenger scale from $10^{5}$ to
$10^{16}$~GeV.  We study various simple modifications of the NMSSM in
Section~5, and find that they either do not break electroweak symmetry
in a phenomenologically viable manner or require a cancellation among
parameters of order $1\%$. 
We finally conclude in Section~6.

\setcounter {equation}{0}

\section{The $\mu$-problem in the GMSB}

In this section, we review the $\mu$-problem in the supersymmetric
Standard Model in general, and also
various attempts to solve it in the context of the GMSB.

The parameter $\mu$ is the only dimensionful quantity present in the
superpotential of the MSSM
\begin{equation}
  W = \mu H_u H_d + \lambda^l_{ij} L_i e_j H_d
  + \lambda^d_{ij} Q_i d_j H_d + \lambda^u_{ij} Q_i u_j H_u.
\end{equation}
Here, $Q_i$, $L_i$, $u_i$, $d_i$, $e_i$ are the matter chiral superfields
with the obvious notation, and $H_u$,
$H_d$ the Higgs doublets.
Note that $\mu$ is part of the supersymmetric Lagrangian,
and hence its origin is, naively, unrelated to the origin of the
soft SUSY-breaking terms
\begin{eqnarray}
  V_{\it soft} &=& m^2_{H_d} |H_d|^2 + m^2_{H_u} |H_u|^2  \nonumber \\
  & & + m_{\tilde{Q}}^{2ij} \tilde{Q}_i^\dagger \tilde{Q}_j 
  + m_{\tilde{L}}^{2ij} \tilde{L}_i^\dagger \tilde{L}_j
  + m_{\tilde{u}}^{2ij} \tilde{u}_i^\dagger \tilde{u}_j +
  m_{\tilde{d}}^{2ij} \tilde{d}_i^\dagger \tilde{d}_j 
  + m_{\tilde{e}}^{2ij} \tilde{e}_i^\dagger \tilde{e}_j \nonumber \\
  & & - m_3^2 H_u H_d
  + {\cal A}_d^{ij} \tilde{Q}_i \tilde{d}_j H_d 
  + {\cal A}_u^{ij} \tilde{Q}_i \tilde{u}_j H_u
  + {\cal A}_l^{ij} \tilde{L}_i \tilde{e}_j H_d.
  \label{eq:soft}
\end{eqnarray}
Phenomenology, on the other hand, dictates that the values of both $\mu$ and
the soft SUSY-breaking masses should be around the weak scale (100
GeV), if SUSY is to be 
responsible for stabilizing the Higgs mass.  Therefore, the important
question is how the mechanism of SUSY breaking can induce a
$\mu$-parameter naturally, at the same order of magnitude as the other
soft SUSY-breaking parameters in the Lagrangian.

One popular scenario of SUSY breaking is the so-called ``hidden
sector'' SUSY breaking in supergravity (SUGRA) \cite{hidden}.  In
hidden sector
models, SUSY is broken in the hidden sector by some mechanism, such
as the
Polonyi model \cite{Polonyi}, gaugino condensation \cite{gaugino},
or the O'Rafeartaigh model \cite{O'Rafeartaigh},
and the effects of SUSY breaking are mediated to the fields in the
supersymmetric Standard Model only by interactions suppressed by the
Planck scale.  It therefore requires SUSY breaking at a scale
$\Lambda \sim 10^{10}$~GeV if the soft SUSY-breaking masses are generated as
$\Lambda^2/M_{\rm Planck}$.  This class of models is able to generate the
appropriate soft SUSY-breaking masses and $\mu$-parameter given that
the $\mu$-term 
is forbidden in the supersymmetric limit by appropriate symmetries,
and arises due to SUSY breaking (see, for example, the Giudice--Masiero
mechanism \cite{SUGRA}).  Hidden sector models
have, on the other hand, to face serious bounds
imposed by flavor-changing neutral currents (FCNC) \cite{DG}.  Low-energy
constraints such as the smallness of $K^0$--$\overline{K^0}$ mixing
require the
matrices $m^{2ij}_{\tilde{Q}}$, $m^{2ij}_{\tilde{d}}$ to have
eigenvalues degenerate to a 
few percent, or their eigenvectors to be strongly
``aligned'' with the eigenvectors of the Yukawa matrices
$\lambda^d_{ij}$ (the same is true for ${\cal A}_d^{ij}$).
 Within the SUGRA
framework alone, there is no
natural mechanism to guarantee the degeneracy or the alignment
\cite{NS}.
In this case, flavor
symmetries are probably necessary to ensure either degeneracy
\cite{non-abelian} or
alignment \cite{NS} and suppress FCNC,
and some of the models presented are also capable of
generating the $\mu$-term through flavor symmetry
breaking \cite{LNS,Nir}.  There is also the possibility that
string theory generates degenerate squark masses if, for instance, the
dilaton field provides the dominant contribution to the soft
SUSY-breaking masses \cite{KL}. 

The gauge mediation of supersymmetry breaking is an alternative mechanism
which can naturally ensure the degeneracy of squarks masses and
therefore suppress the
dangerous FCNC effects.  SUSY is somehow broken 
(hopefully dynamically via dimensional transmutation to generate a
large hierarchy), and SUSY-breaking effects are mediated to the
fields in the supersymmetric Standard Model by the Standard Model
gauge interactions.  Mediating SUSY breaking via gauge interactions is
not a novel idea \cite{SUSY3,oldGM}. It
allows for SUSY breaking at a lower scale (when compared to SUGRA inspired
models) and, because all SUSY-breaking effects are
transmitted by flavor blind interactions (the Standard Model
gauge interactions), squarks of different families have the same
mass. This scheme has attracted a lot of interest after the
pioneering works by the authors of references \cite{DN,DNS,DNNS}, which showed
that one can
successfully mediate the SUSY-breaking effects via gauge interactions
with the help of a so-called ``messenger sector.''  Their scheme can easily
incorporate dynamical SUSY breaking and can explain the origin of the
large hierarchy between the Planck (string, grand unified (GUT)) scale
and the weak scale.

The GMSB itself, however,  has nothing
to say about the $\mu$-parameter unless one
introduces extra fields which couple to the particle content of the MSSM.
The $\mu$-problem in the GMSB is the primary interest of
this paper.
Many solutions to the $\mu$-problem have been suggested by different authors
and all of them require the introduction of new fields and/or interactions.
Some of these solutions will be reviewed shortly.

In the original models \cite{DN,DNS,DNNS}, SUSY is broken
dynamically in a so-called SUSY-breaking sector and the breaking
effects are transmitted to the
supersymmetric Standard Model via a messenger sector.  The energy
scale of the messenger sector is given by $\Lambda \simeq
10^4$--$10^5$~GeV.  There are,
however, models which do not have a separate messenger sector so that
the sector which breaks SUSY dynamically is directly coupled to the
Standard Model gauge group \cite{direct, Amess,direct2}.  In this
case, the effective messenger scale tends to be much higher.
For our purposes it is enough to employ a simple version of the
messenger sector, as in the original models, and take the messenger
scale $\Lambda$ as a free parameter.  

The messenger sector can be described by the superpotential
\begin{equation}
  \label{messengerW}
  W_{S}={1\over 3}\lambda S^{3}+ \kappa S \Phi^{+}\Phi^{-}
  + \kappa_{q}Sq\bar q + \kappa_{l}S l\bar l,
\end{equation}
where $S$ is a singlet superfield, $\Phi^{\pm}$ are charged under a $U(1)$
associated with the SUSY-breaking sector and are singlets under the
Standard Model $SU(3)\times SU(2)\times U(1)$ gauge group.
The superfield $q$ ($\bar q$) transforms as a 
$({\bf 3} ({\bf\bar 3}),{\bf 1},\pm 1/3)$
under the
Standard Model, while $l$ ($\bar l$) transforms as $({\bf 1},{\bf 2},\mp
1/2)$.

We assume that
the scalar components of $\Phi^{\pm}$ acquire negative SUSY-breaking
masses-squared
due to its interaction with the SUSY-breaking sector (usually
accomplished by the so-called ``messenger $U(1)$'' gauge interaction
\cite{DNS,DNNS}), and
the potential associated with the scalar component of $S$ reads
\begin{equation}
  V_{S}=-| m_{\Phi}|^{2} ( | \Phi^{+}|^{2} + | \Phi^{-}|^{2})
  + | \kappa S \Phi^{+}|^{2} + | \kappa S \Phi^{-}|^{2}
  + | \kappa \Phi^{+}\Phi^{-}+\lambda S^{2}| ^{2},
\end{equation}
neglecting terms containing $l$ or $q$. It is easy to see that the
scalar and the
$F$ components of $S$ acquire vacuum expectation values (VEVs)
$\langle S \rangle$
and  $\langle F_{S}\rangle$ and therefore $q$ and $l$ acquire supersymmetric
masses proportional to $\langle S\rangle$ and SUSY-breaking
masses-squared proportional to 
$\langle F_{S}\rangle$.\footnote{There is a run-away direction
  $q=\bar{q}$, $l=\bar{l}$ in this potential \cite{DDRandall}.  This
  problem can be
  avoided by introducing more $S$ fields to the messenger sector.
  Such details are, however, irrelevant for the rest of our
  discussion.}   This effect feeds down to the
MSSM through loop corrections. Gauginos acquire Majorana masses at one loop,
while sfermions acquire SUSY-breaking masses-squared at two loops. The
calculation of these soft SUSY-breaking parameters was done long ago
(see \cite{SUSY3,oldGM}) 
 and its result is well known. At the messenger scale:
\begin{equation}
\label{gauginomass}
M_{i}={\alpha _{i}\over 4\pi}nB,
\end{equation}
\begin{equation}
\label{sfermionmass}
m^{2}_{\tilde f_{i}}=2nB^{2}\left({3\over
 5}Y_{i}^{2}\left({\alpha_{1}\over 4\pi}\right)^{2} 
 + C_{2i}\left({\alpha_{2}\over 4\pi}\right)^{2}
+C_{3i}\left({\alpha_{3}\over 4\pi}\right)^{2}\right).
\end{equation}
Here and below, all $\alpha_i = g_i^2/4\pi$ are in the $SU(5)$
normalization, $B = \langle F_{S}\rangle/\langle S\rangle$ in the
messenger sector discussed above,  and
$n$ determines the number of messenger sector superfields responsible for
mediating SUSY breaking. In the example we described above, which will be
referred to as the model with the minimal GMSB, $n=1$. $Y$ is the
hypercharge of the particle, $C_{2}=3/4$ for weak $SU(2)$ doublets
(zero for singlets) and 
$C_{3}=4/3$ for color triplets (zero for color singlets).
Eq.~(\ref{sfermionmass}) guarantees that squarks of different
families are degenerate at the messenger scale and therefore FCNC
effects are safely suppressed.
It is interesting to note that, for small $n$, gaugino masses and sfermion
masses are comparable. For very large $n$, on the other hand, sfermion
masses can be significantly smaller than gaugino masses (by a factor $\sqrt
n$).

In the mechanism described above, trilinear couplings are not
generated at the same order (in loop expansion)
at the messenger scale. This is not the case in general, and some models
can generate trilinear couplings with values comparable to the other
soft SUSY-breaking parameters even at
the messenger scale \cite{Amess}. We will, for most of our discussions,
 consider
\begin{equation}
\label{Aterms}
{\cal A}^{ij}_{l,u,d}(\Lambda)=0,
\end{equation}
unless otherwise noted.

The GMSB does not generate a
$\mu$-term because of the non-renormalization theorem. Therefore $\mu$
is an input of the model, and, because it 
has dimensions of mass, its only nonzero natural value 
is $M_{Planck}$ ($M_{string}$, $M_{GUT}$). This is clearly
not allowed phenomenologically.
The $\mu$-term must, therefore, be forbidden at the Planck scale (by, say, a
$Z_{3}$ symmetry) and generated dynamically.  Below, we review various
attempts to generate the $\mu$-term in the context of the GMSB.  The
following list is not meant to be exhaustive and our descriptions of
the various attempts are by no means complete.
The review below only intends to show that many attempts have been
made while none of them appears to be entirely satisfactory.  

The simplest solution would be to introduce a term in the
superpotential \cite{DNS}
\begin{equation}
\label{simple}
W\supset k S H_{d} H_{u},
\end{equation}
where $S$ is the singlet superfield in Eq.~(\ref{messengerW}). In such a
scenario $\mu = k\langle S\rangle$ and $m_{3}^{2}=k
\langle F_{S}\rangle$.
$m_{3}^{2}$ is the SUSY-breaking Higgs mixing mass-squared in
Eq.~(\ref{eq:soft}).

Phenomenology imposes that both $\mu$ and $\sqrt{m_{3}^{2}}$ are of the order
of the weak scale, unless one is willing to accept a drastic
cancellation among parameters to reproduce the observed $M_Z$. Therefore,
\begin{equation}
\label{mu=Bmu}
(k\langle S\rangle)^2\sim k\langle F_{S}\rangle\sim (100~\mbox{GeV})^{2},
\end{equation}
\begin{equation}
\label{FS/M=100}
{\langle F_{S}\rangle\over \langle S\rangle}\sim k\langle S\rangle
\sim 100~\mbox{GeV}
\end{equation}
and
\begin{equation}
\label{muBmu}
m_{3}^{2}=\mu {\langle F_{S}\rangle\over \langle S\rangle}.
\end{equation}

This situation is already excluded
experimentally. Eq.~(\ref{gauginomass}) states
that the gluino mass is given by $(\alpha_{3}/4\pi)\langle F_{S}\rangle/
\langle S\rangle$,
and if Eq.~(\ref{FS/M=100}) is satisfied one would arrive at $M_{\tilde
  g}\simeq1$~GeV, which is unacceptable. The same is true for all the
other soft SUSY-breaking masses.
This is a general consequence of Eq.~(\ref{muBmu}). It implies that
$\sqrt{m_{3}^{2}}\gg \mu$ if all experimental bounds on the SUSY spectrum are
to be satisfied, while $SU(2)\times U(1)$ breaking requires
Eq.~(\ref{mu=Bmu}). Some authors refer to this puzzle as the $\mu$-problem
in the GMSB \cite{DGP}.

Another simple solution that does not require the introduction of any
extra superfields into the theory couples the Higgs superfields to the
$q$ superfields present in Eq.~(\ref{messengerW}) \cite{DGP}.
In the minimal messenger sector \cite{DNS,DNNS}, one may have, instead
of $q$ and $l$, a complete ${\bf
5}+{\bf\bar 5}$ multiplet of $SU(5)$ to preserve the gauge coupling
unification.  One can also use a ${\bf 10} + {\bf \overline{10}}$ for this
purpose, and generate gaugino masses and scalar masses-squared with $n=3$.
In this case, one can couple the components $Q$
in {\bf 10} that have the same quantum numbers as left-handed quark
doublets and the components $u$ that have the same quantum numbers as
right-handed up quarks (or their corresponding
components in ${\bf \overline{10}}$) to the Higgs doublets. 
Explicitly, $W\supset
\lambda_{1} H_{d} \bar{Q} \bar{u} +\lambda_{2} H_{u}
Q u$.  This will
induce, in the Lagrangian, a one-loop term proportional to
\begin{equation}
{\lambda_{1}\lambda_{2}\over16\pi^{2}}\int d^{4}\theta
{H_{d}H_{u}S^{\dagger} S^{\dagger}\over S^{\dagger}S} \; .
\end{equation}
The $F$ vacuum expectation value of $S$ will generate
$\mu\simeq{\lambda_{1}\lambda_{2}\over16\pi^{2}}{\langle F_{S}\rangle\over
\langle S\rangle}$
and $m_{3}^{2}\simeq{\lambda_{1}\lambda_{2}\over16\pi^{2}}\left(
{\langle F_{S}\rangle\over \langle S\rangle}\right)^2$.
Again one runs into Eq.~(\ref{muBmu}) and must hunt for other solutions.

All of the models described above couple the MSSM Higgs superfields to
those in the messenger sector. Not only did we encounter the problem of
Eq.~(\ref{muBmu}), but some of the coupling constants introduced had to
be made fairly small because of the magnitude of $\langle S\rangle$ and
$\langle F_{S}\rangle$. Another class
of solutions tries to get around this issue by introducing another singlet
superfield, whose vacuum
expectation value would generate the $\mu$-term.

One motivation for such models is to utilize the extra singlet to solve
the doublet-triplet Higgs splitting in $SU(5)$ grand unified theories via
a sliding singlet mechanism \cite{sliding}.  This mechanism is known
to be unstable
against radiative corrections if the soft SUSY-breaking parameters are
generated 
at a scale higher than the GUT scale, but can be stable for the
low-energy GMSB \cite{Nemeschansky}.
Ciafaloni and Pomarol \cite{CP} claim
that such a solution would generate a viable
$\mu$-term. We believe, however, that the conditions that they impose on the
soft SUSY-breaking parameters can never be satisfied
in the context of the GMSB, where all soft SUSY-breaking masses
are tightly related.  We will comment on this in Section~\ref{sec:mod}.

The simplest model with the addition of an extra singlet one can
imagine, referred to as the NMSSM \cite{Eetal}, involves substituting
the $\mu$-term in the MSSM
superpotential by
\begin{equation}
\lambda H_{d}H_{u}N-{k\over 3}N^3.
\end{equation}
The minimization
of the scalar potential for $H_{d}$, $H_{u}$ and $N$
at the weak scale
should produce VEVs $v_{d}$ and $v_{u}$ for both Higgs bosons, thus breaking
$SU(2)\times U(1)$, and $x$ for the singlet. $\mu$ would be  equal to
$\lambda x$. The $m_{3}^{2}$ term would arise due to renormalization group
(RG) running of the $A$-term
$\lambda A_{\lambda}H_{d}H_{u}N$ from the messenger scale to the
weak scale. $m_{3}^{2}$ would be equal to
$\lambda A_{\lambda} x$.

Dine and Nelson \cite{DN} claim that this model does not work for the
low-energy GMSB.  A detailed analysis was not presented in
their paper, and we will explain the problem in Section~\ref{sec:NMSSM}.
They suggest the introduction of an extra light pair of $q' + \bar q'$
and $l' + \bar l'$ as a means to produce a viable spectrum.  They did not,
however, publish a quantitative analysis of the model, and say nothing
about its
naturalness.  Agashe and Graesser \cite{AG} study this scenario
and show that there is indeed a solution, but it is fine-tuned.  They
present a possibility to ease the fine-tuning by employing many
lepton-like messengers while keeping the number of quark-like
messengers small.  
In Sections~\ref{sec:NMSSM} and \ref{sec:mod}, we analyze in
great detail the case for
both the high- and low-energy GMSB.

There are ways of giving $N$ a VEV which are not related to
electroweak symmetry breaking. In Ref.~\cite{DNS} two mechanisms
are introduced, neither of them very appealing, where the $N$ VEV is
generated at the messenger scale. Namely,
\begin{equation}
\label{DNS1}
W\supset -\frac{1}{2} k_{S} N^2 S
\end{equation}
or
\begin{equation}
\label{DNS2}
W\supset -k_{q}Nq\bar q-k_{l}Nl\bar l
\end{equation}
in addition to the NMSSM.  $S$, $q$ and $l$ are the messenger sector
superfields present in Eq.~(\ref{messengerW}).

In the case of Eq.~(\ref{DNS1}), a potential
\begin{equation}
V_{N}=| kN^{2}+k_{S}\langle S\rangle N |^{2}-k_{S}N^2\langle F_{S}\rangle
\end{equation}
is generated for $N$ in the presence of $\langle S \rangle$ and
$\langle F_{S} \rangle$ VEVs. If one assumes $k_{S}$ to be small,
$N$ develops a VEV $x=\sqrt{{k_{S}\langle F_{S}\rangle\over k^2}}$,
and $\mu={\lambda\over k}\sqrt{k_{S}\langle F_{S}\rangle}$ assuming all other
couplings to be of order one. It is easy to see, a posteriori, that
$k_{S}$ must indeed be small if one is to generate a
phenomenologically viable $\mu$.  Unfortunately this case requires
that the soft SUSY-breaking masses-squared $\sim (\alpha_{i}/4\pi)^{2} (\langle
F_{S}\rangle/\langle S \rangle)^{2}$ and $\mu^{2} \sim k_{S}\langle
F_{S}\rangle$ are accidentally of the same order of
magnitude.\footnote{Although $k_{S}$ has to be small, its
smallness is natural in the sense of 't Hooft. It can be
interpreted as being generated due to the breaking of some global
symmetry, such as $N\rightarrow e^{2\pi i/ 3}N$ and
$H_{1,2}\rightarrow e^{2\pi i/ 3}H_{1,2}$, while $S$ is
invariant. This type of symmetry would also explain the suppression
of a term $NS^2$ in the superpotential, which would be of order
$(k_{S})^2$.}

The superpotential coupling Eq.~(\ref{DNS2}) would lead to a potential
\begin{equation}
V_{N}\supset | kN^{2}|^{2} - k_{q}N{1\over 32 \pi^2}
{(\kappa_{q}\langle F_{S}\rangle)^2\over \kappa_{q}\langle S\rangle} -
(l \leftrightarrow q).
\end{equation}
The linear terms in $N$ arise via tadpole one-loop diagrams involving $q$'s and
$l$'s. This would lead to $x^3={k_{q}\over k^{2}}
{1\over 32 \pi^2}{\kappa_{q}\langle F_{S}\rangle^2\over
\langle S\rangle}+(l\leftrightarrow q)$. Again $k_{q}$
and $k_{l}$ would have to be small. This solution
still faces the problem of explaining why a term $NS^{2}$ is
not present in the superpotential. Note that the presence of such a
term would lead
to an unacceptably large VEV for $N$.  One may argue, however, that this is
``technically natural'' because the absence of a term in the
superpotential is preserved by radiative corrections.  An even more
serious problem is the need to suppress the kinetic mixing $\int
d^{4} \theta S^{\dagger} N + h.c.$ to ensure $F_{N} \ll F_{S}$;
an unacceptably large $m_{3}^{2} = \lambda F_{N}$ would be generated
otherwise.
An order unity kinetic mixing can be induced
via radiative corrections between the ultraviolet cutoff, say
the Planck scale, and the messenger scale, and the bare parameter
has to be chosen very carefully so that
the unwanted mixing term can be canceled at the messenger scale.
This kinetic mixing can be forbidden if there are two sets of
messenger fields and if the field $N$ couples off-diagonally, {\it
  e.g.}\/, $W = N q_1 \bar{q}_2$ etc \cite{GR}.
Then the tadpole term mentioned
above is also forbidden, but a negative mass squared for the $N$ field
can be generated instead.  This would lead to the NMSSM in a
successful manner; again the parameters must be carefully
chosen as in the NMSSM with extra light quark pairs (see Section~5).

Another solution with extra singlets, which points an interesting way around 
Eq.~(\ref{muBmu}), was suggested by Dvali, Giudice and Pomarol \cite{DGP}.
Their idea is to generate the $\mu$-term via the following one-loop 
effective term in the Lagrangian:
\begin{equation}
\label{dgpddr}
\int d^{4}\theta {H_{d}H_{u}D^{\alpha}D_{\alpha}(S^{\dagger}S)\over
\langle S\rangle^3},
\end{equation}
where $D_{\alpha}$ is the supersymmetric covariant derivative.
This works because $D^2$ cancels $\theta^2$
in $S$, while leaving $\bar{\theta}^2$ in $S^\dagger$.  Then the
integral over $d\bar{\theta}^2$ can be done and the $\mu$-term is
generated, while $m_3^2$ is not.  The $m_{3}^{2}$ term
would arise at higher loops, or via some other mechanism.

An explicit realization of this mechanism \cite{DGP} is the following.
Suppose a singlet field $N$ acquires a linear term $M^2 N$ in the
superpotential due to its coupling to the messenger sector.  Then the
superpotential $W = N(Y^2 + H_u H_d - M^2)$ leads to a minimum with
$\langle Y\rangle=M$, $N=0$. However, by further coupling $N$ to the
messenger superfields, {\it i.e.}\/ $N q\bar{q}$ etc, a one-loop diagram 
of messenger
fields generates the operator $\frac{1}{16\pi^2} \int d^4 \theta N
S^\dagger (S^\dagger S)/\langle S^\dagger S\rangle$, which contains $V
\sim \frac{1}{16\pi^2} N \langle F_S\rangle^2/\langle S\rangle$.
Note that this is the same linear potential generated in the case of 
Eq.~(\ref{DNS2}).  This
tadpole term induces a VEV for $N$ of order $\langle N \rangle \sim
\frac{1}{16\pi^2}\langle F_S\rangle^2/\langle S\rangle/\langle Y\rangle^2$ 
which
is of the order of the weak scale if $\langle Y\rangle^2=M^2 \sim 
\langle F_S \rangle$.
The $Y$ field plays a crucial role: it
{\it slides}\/ 
to cancel the
$F$-component VEV of $N$ before the tadpole is added and, after SUSY is 
broken, its VEV is shifted and leads to $\langle F_{N}\rangle=m_3^2\sim 
\mu^2$, as required by phenomenology. Note that the $\mu$-parameter
obtained here can be understood as a consequence of the effective Lagrangian
Eq.~(\ref{dgpddr}), which is generated  upon integrating out $N$ and
$Y$ before substituting the effect of the VEVs of $S$.
 
The necessary linear term ($M^2N$) in the superpotential for $N$ can
be easily generated by the kinetic mixing between $N$ and $S$ 
or also by other mechanisms, as pointed out in reference \cite{DDR}.
One apparent drawback of this realization is that one needs a set
of new fields whose interactions are arranged in a rather special way. 
Furthermore one would expect the presence of a term proportional to
$SH_uH_d$ in the superpotential. This happens because both $S$ and $N$
couple to the messengers, that is, $W\supset Sq{\bar q}+Nq{\bar
  q}$, and have, therefore, the same quantum numbers. 
We have already argued that
a coupling $SH_uH_d$ has to vanish (see Eq.~(\ref{simple})).  
Finally we point out that this model also suffers from the
cancellation problem present in the MSSM (see Section 3). 


Dine, Nelson, Nir and Shirman \cite{DNNS} suggest
yet another way of generating a $\mu$-term with the introduction of an extra
singlet.  It was inspired by flavor symmetry models in \cite{LNS},
and resembles a modified version of the NMSSM + Eq.~(\ref{DNS1}):
\begin{equation}
\label{flavorinsp}
W\supset \lambda_n {N^{n+1}\over M_{Pl}^{n}}H_{d}H_{u}
+\lambda_m{N^{m+3}\over M_{Pl}^{m}}+
\lambda_p{N^{2+p}\over M_{Pl}^{p}}S
\end{equation}
where $M_{Pl}$ is the Planck mass. When $m=2$, $n=1$ and $p=2$,
it is easy to check that $\mu\sim\lambda_n\sqrt{\langle
  F_{S}\rangle}$. We assume the other couplings to be of order 1. It is also 
easy to see that one would require a very small, carefully chosen
coupling $\lambda_n$ in order to guarantee $\mu\sim 100$~GeV. It is
worth noting that  
this mechanism does not generate an $m_{3}^{2}$ term.

At last we would like to mention another interesting possibility,
pointed out by Yanagida \cite{Y} and Nilles and Polonsky \cite{NP}.  Their
models utilize the accidental equality 
$(\Lambda_{\rm DSB}/M_{*})^{1/3} \sim (\alpha/4\pi)^2$, where
$\Lambda_{\rm DSB} \sim 10^7$~GeV is the scale of dynamical SUSY
breaking (DSB) in models with
the low-energy GMSB and $M_* = M_{Pl}/\sqrt{8\pi}$ the reduced Planck
mass.  By introducing a new SUSY-preserving sector  
with strong gauge dynamics, Yanagida's model generates a VEV for the
superpotential 
which cancels the cosmological constant from the DSB sector.  The
constant superpotential in turn generates a $\mu$-term of order
$\Lambda_{\rm DSB}(\Lambda_{\rm DSB}/M_{*})^{1/3} \sim
(\alpha/4\pi)^2 
\Lambda_{\rm DSB} \sim 1$~TeV.  The phenomenology of this
model is the same as the previous one (see Eq.~(\ref{flavorinsp})).
The model by Nilles and Polonsky makes use of the Planck-scale suppressed
K\"ahler potential, $\int d^4\theta N (z^* z) /M_{*}$,
where $z$ is a 
chiral superfield in the DSB sector with an $F$-component VEV.  This
operator may be present at the tree-level, but may also be generated
by gravitational effects.  It generates a tadpole for the singlet
$N$: $V = (\Lambda_{\rm DSB}^4/M_*) N$.  Together with the ${k\over 3} N^3$
superpotential of the NMSSM, it generates a VEV for $N$ of order
$\langle N \rangle \sim (\Lambda_{\rm DSB}^4/M_*))^{1/3}$.  Even
though these models generate the correct $\mu$-term of order the weak
scale in the models with the low-energy GMSB, this would not work for the
high-energy GMSB.  

We consider that none of the mechanisms outlined above are entirely
satisfactory. Most of them require a very specific choice of parameters and
the introduction of extra matter at or slightly above the weak scale.
Furthermore, most of them have not been studied quantitatively (see,
however, Ref.~\cite{Sarid}), and there is
no guarantee that they indeed generate the correct electroweak symmetry
breaking pattern and an experimentally viable spectrum. And last,
but not least, there is no study of how natural such a solution is,
given that a viable pattern of electroweak symmetry breaking can be 
generated.

It is, therefore, part of our goal to study the simplest of the
models mentioned above in detail. Before that, we would like to
review the status of electroweak symmetry breaking in the MSSM,
where the $\mu$-term is introduced ``by hand.'' We will point out
that, in the case of the GMSB, the current
lower bounds on superparticle masses already require an order $10$\%
cancellation 
between the $\mu$-parameter and the soft SUSY-breaking parameters.

\setcounter {equation}{0}
\section{The $\mu$-parameter in the MSSM}
\label{mssmsection}

We reviewed various proposals to generate the $\mu$-parameter in
models with the GMSB.
In this section, we review how electroweak symmetry breaking
occurs in the MSSM, assuming that the $\mu$-parameter and $m_{3}^{2}$ are
somehow generated.  In particular, we point out a need for an order $10$\%
cancellation between $\mu$-parameter and soft SUSY-breaking parameters
in models 
with the GMSB given the current experimental lower bounds on superparticle
masses.  Note that the case of the NMSSM is different because the
$\mu$-parameter is generated together with electroweak symmetry breaking and
hence the two problems cannot be clearly separated.  This will be
discussed in the next two sections.

The tree-level Higgs potential in the MSSM is given by
\begin{equation}
        V = m_{1}^{2} |H_{d}|^{2} + m_{2}^{2} |H_{u}|^{2}
                - m_{3}^{2} (H_{d} H_{u} + c.c.)
                + \frac{g_{2}^{2}}{8} (H_{d}^{\dagger} \vec{\sigma} H_{d}
                        + H_{u}^{\dagger} \vec{\sigma} H_{u}) ^{2}
                + \frac{g'^{2}}{8} (|H_{d}|^{2} - |H_{u}|^{2})^{2},
\end{equation}
where the mass parameters involve both the supersymmetric $\mu$-term and
the soft SUSY-breaking terms,
\begin{eqnarray}
        m_{1}^{2} & = & \mu^{2} + m^{2}_{H_{d}},
          \\
        m_{2}^{2} & = & \mu^{2} + m^{2}_{H_{u}}.
\end{eqnarray}
In the MSSM, one can show that the vacuum can always be gauge rotated to
the following configuration
\begin{equation}
        H_{d} = \left( \begin{array}{c} v_{d} \\ 0 \end{array} \right),
        \qquad
        H_{u} = \left( \begin{array}{c} 0 \\ v_{u} \end{array} \right).
\end{equation}
The two expectation values need to satisfy $v_{d}^{2} + v_{u}^{2} =
v^{2} = (174~\mbox{GeV})^{2}$ in order to reproduce the observed
$M_{Z}$, and it is conventional to parametrize
them by $v_{d} = v \cos\beta$, $v_{u} = v \sin\beta$.
The minimization condition of the potential can be rewritten in the
following form:
\begin{eqnarray}
        \frac{M_{Z}^{2}}{2} & = & -\mu^{2} +
                \frac{m_{H_{d}}^{2} - m_{H_{u}}^{2} \tan^{2}\beta}
                {\tan^{2} \beta -1},
                \label{eq:mueq}\\
        2 m_3^2&  =  & (2 \mu^{2} + m_{H_{d}}^{2} + m_{H_{u}}^{2})\sin 2\beta .
        \label{eq:m3eq}
\end{eqnarray}

It has been claimed that electroweak symmetry breaking is natural
in the MSSM because $m^{2}_{H_{u}}$ is easily driven negative due to
the presence
the top Yukawa coupling in its RG evolution.  In
models with the minimal GMSB such as the original ones
in \cite{DNS,DNNS}, the boundary
condition for the supersymmetry breaking parameters are given by
Eqs.~(\ref{gauginomass}, \ref{sfermionmass}, \ref{Aterms}).
A simple one-loop approximation is valid in the case of the low-energy
GMSB  because of the small logarithm between the messenger scale
$\Lambda$ and
the electroweak scale, and one finds
\begin{equation}
        m^{2}_{H_{u}} ( M_{Z}) \simeq
                m^{2}_{H_{d}} (M_{Z})
                - \frac{6}{16\pi^{2}} h_{t}^{2} 2 m_{\tilde{t}}^{2}
                \log \frac{\Lambda}{M_{Z}},
\end{equation}
which is always negative.

The need for a cancellation between the $\mu$-parameter and soft
SUSY-breaking masses can be seen as follows.
Experimental constraints bound the superparticle masses from below,
which hence set a lower limit for the ratio
$B=\langle F_{S}\rangle /\langle S
\rangle$.  Therefore one finds that $|m^{2}_{H_{u}}|$ is
bounded from below.  On the other hand, in order for the observed
$M_{Z}^{2}$ to be reproduced, the $\mu$-parameter is constrained by
Eq.~(\ref{eq:mueq}).
For a moderately large $\tan\beta \gtrsim 2$, $m^{2}_{H_{d}}$ can be
completely neglected and one finds
\begin{equation}
        \frac{M_{Z}^{2}}{2} \sim - \mu^{2} - m^{2}_{H_{u}}.
\end{equation}
This equation requires a cancellation between $\mu^{2}$ and
(negative) $m^{2}_{H_{u}}$ to reproduce $M_{Z}^{2}/2 \sim
(70~\mbox{GeV})^{2}$ correctly.  The degree of
cancellation is given by $(M_Z^2/2)/\mu^2$.\footnote{The degree
of cancellation is defined as follows: it is a percentage
quantity that measures how much a given input parameter
(in this case $\mu ^{2}$)
is free to vary before a given output parameter (in this case $M_{Z}^{2}$)
changes significantly.  Explicitly, the degree of cancellation is
$({\rm d}(\log M_{Z}^{2})/ {\rm d}(\log\mu ^2))^{-1}$.  This definition
corresponds to the inverse of the Barbieri--Giudice function \cite{BG}.}

To determine the lower limit on $|m^{2}_{H_{u}}|$, we consider a
number of
experimental constraints \cite{Jerusalem}.  One is that the gluino
must be heavier than 
190~GeV, which becomes stronger if the squarks have
comparable masses.  The second is that the right-handed selectron
must be heavier than 80~GeV.\footnote{This bound depends on the mass
  of the neutralino into which the selectron decays.  However, since $\mu$
  turns out to be large, it is {\it a posteriori}\/ justified to
  assume that the lightest neutralino is almost pure bino.  Then the GMSB
  predicts the relation between selectron and the bino masses,
  and hence we have a fairly reliable lower bound.}   For large
$\tan\beta$, the right-handed stau may become rather light; we then require
$m_{\tilde{\tau}} > 55$~GeV if it decays into tau and a neutralino
or gravitino, and $m_{\tilde{\tau}} > 73$~GeV if it does not decay
inside the detector. We also considered the lightest chargino to be
heavier than 63 GeV. The most recent lower bound on the chargino mass
\cite{LEP2} is $m_{\chi^{+}}\gtrsim 67$~GeV, which leaves our analysis
virtually unchanged.

Let us first discuss the case of the minimal low-energy GMSB
with small $\tan\beta$ to make the argument clear.  Here we consider
$\Lambda\sim 10^5$~GeV.  The gluino mass
constraint requires
\begin{equation}
        B > 23~\mbox{TeV}.
\end{equation}
This bound itself is independent from the messenger scale.  However,
the gluino mass bound depends on the mass of the squarks, and it
strengthens if the squark masses are comparable to the gluino mass.  For
the minimal low-energy GMSB, squarks are significantly heavier than
the gluino and we can use the bound above.
A more stringent constraint is derived from the requirement that the
right-handed sleptons are heavier than 80~GeV.  Including the
one-loop renormalization group evolution and the $D$-term, we find
\begin{eqnarray}
        m_{\tilde{e}}^{2} &=& 2 \frac{3}{5}
        \left(\frac{\alpha_{1}}{4\pi}\right)^{2} B^2
        + \frac{2}{11} \left(\left(\frac{\alpha_{1}(M_{\it
        mess})}{\alpha_{1}(M_{Z})}\right)^{2}-1\right) M_{1}^{2}(M_{Z})
        + M_{Z}^{2} \sin^{2}\theta_{W} \cos 2\beta \nonumber \\
        &=& 2.89\times 10^{-6}B^2
        - 0.232 M_{Z}^{2} \cos 2\beta.
\end{eqnarray}
Therefore we find
\begin{equation}
        B > 39~\mbox{TeV}
\end{equation}
for the most conservative case $\cos 2\beta = -1$.  With this lower
bound we find
\begin{equation}
        m^{2}_{\tilde{t}} > 2 \frac{4}{3}
        \left(\frac{\alpha_{3}(M_{\it mess})}{4\pi}\right)^{2} B^2
        > (430~\mbox{GeV})^2 .
\end{equation}
Using the one-loop running of $m^{2}_{H_{u}}$, we obtain
\begin{equation}
        m^{2}_{H_{u}} (M_{Z}) < - (260~\mbox{GeV})^{2}
\end{equation}
and as a result of the minimization condition,
\begin{equation}
        \mu > 250~\mbox{GeV}.
\end{equation}
This requires a cancellation of 7\% in order to obtain the correct
$M_{Z}^{2}$.  Even though this level of
cancellation is not of immediate concern, this analysis shows the
need for a certain amount of cancellation which will become worse as
experimental lower bounds on superparticle masses improve.

\begin{figure}[tb]
  \begin{center}
    \leavevmode
    \epsfxsize=0.6\textwidth \epsfbox{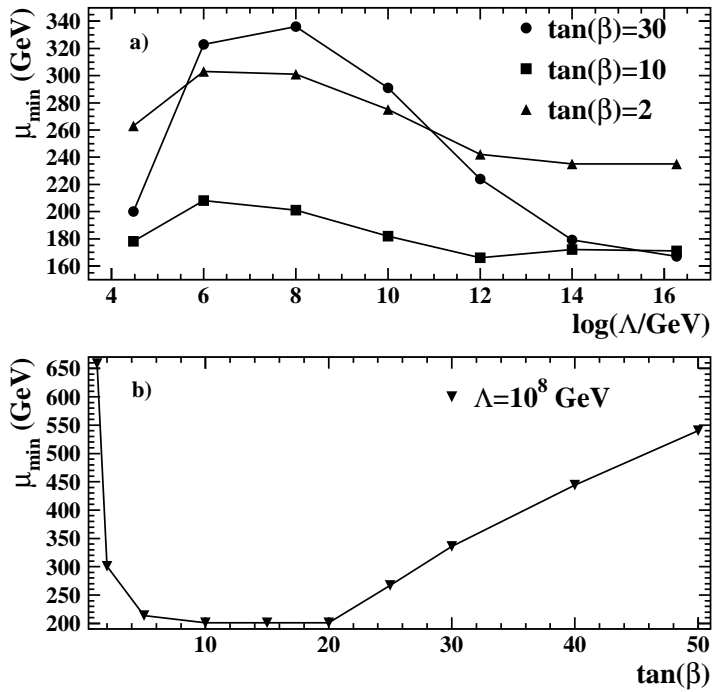}
    \caption[mumin]{Lower bounds on $\mu$ in models with the GMSB
      subject to the constraint $M_{Z}=91$~GeV and to the lower 
      bounds on superparticle masses (see text), (a) as a
      function of the messenger scale, for $\tan\beta=$ 2, 10, and 30,
      and (b) as a
      function of $\tan\beta$ for a fixed messenger scale of $10^8$~GeV.}
    \label{fig:mumin}
  \end{center}
\end{figure}

As it is clear from the  argument above, the actual lower bound on
$\mu$ depends on the messenger scale and $\tan\beta$.  We have
studied this issue numerically  using the experimental
bounds quoted above and found the lowest possible value of $\mu$ as a
function of
the messenger scale. In Fig.~\ref{fig:mumin}(a) we present bounds for three
values of $\tan\beta$.  The lower bound on $\mu$ comes from one of the
various
experimental constraints.  For instance, the $\tan\beta=2$ case is
dominated by the lower bound on $m_{\tilde{e}_R}$ up to a messenger
scale of $10^{12}$~GeV, after which the gluino mass bound is more
important.\footnote{We only analyzed the case for one messenger
  ($n$=1). For larger $n$ the gluino bound becomes less
important and the slepton bounds dominate up to the GUT scale.}
The case of $\tan\beta=10$ has a similar
behavior.  The
situation is more complex and interesting for $\tan\beta=30$.  For a
messenger scale of up to $10^{10}$~GeV, the stau is the lightest
supersymmetric particle (except for the gravitino).  It decays inside the
detector to tau and gravitino for the lowest messenger scale, but
leaves the detector without decaying for higher messenger scales.
This stable stau provides the strongest constraint.  From messenger
scales above $\sim 10^{12}$~GeV the stau decays inside the detector
to tau and neutralino. This bound dominates up to $\sim
10^{16}$~GeV, when the
gluino bound dominates. The chargino bound is
comparable to that of the gluino for the GUT scale
($M_{GUT}=1.86\times 10^{16}$~GeV).

In Fig.~\ref{fig:mumin}(b) we show the minimum value of $\mu$ as a function of
$\tan\beta$ for a fixed messenger scale ($\Lambda = 10^{8}$~GeV).  The
$\tan\beta$ dependence can be easily understood as follows.  Starting
from low $\tan\beta$, increasing $\tan\beta$ decreases the top Yukawa
coupling, and hence $m_{H_u}^2$ receives a less negative contribution
from the top-stop loop.  Therefore a lower value of $\mu$ is allowed.
This part is dominated by the $\tilde{e}_R$ bound.
However beyond $\tan\beta \sim 20$, the bottom and tau Yukawa coupling
become important.  In fact, the scalar tau mass is pushed down both
because of the loop effect and left-right mixing, and the experimental
lower bound on $B$ becomes stronger.  Beyond $\tan\beta \sim 30$,
the stau does not decay inside the detector for this choice of
the messenger scale and the constraint is even more stringent.  
This in turn leads to
a more negative $m^2_{H_u}$ and hence a larger $\mu$.  

Combining both the
messenger scale dependence and $\tan\beta$ dependence, we conclude
that the most conservative current limit is
\begin{equation}
  \mu > 160~\mbox{GeV}.
\end{equation}
The required cancellation between $\mu^{2}$ and soft SUSY-breaking
parameters in 
order to reproduce the observed $M_{Z}$ is $M_{Z}^{2}/2\mu^{2} =
16$\%.  Note that this level of cancellation is the absolute minimum,
and a more accurate cancellation is required for most of the
parameter space.

In the case of minimal supergravity models, where all scalars have the
universal SUSY-breaking mass-squared $m_{0}^{2}$, all gauginos have mass
$M_{1/2}$ and all $A$-terms are given by ${\cal A}_f^{ij} =
A_{0}\lambda_f^{ij}$ for $f=u,d,l$, at the
GUT scale, the situation appears to be much
better.  The renormalization group equations can be solved numerically
for each choice of $\tan\beta$.  As an example we take $\tan\beta=2$ and
find
\begin{eqnarray}
        m^{2}_{H_{1}} & = & m_{0}^{2} + 0.50 M_{1/2}^{2},
          \\
        m^{2}_{H_{2}} & = & -0.32 m_{0}^{2} - 2.49 M_{1/2}^{2} - 0.05 A_{0}^{2}
                -0.20 M_{1/2} A_{0},
          \\
        m^{2}_{\tilde{e}} & = & m_{0}^{2} + 0.15 M_{1/2}^{2} .
\end{eqnarray}
By requiring $m_{\tilde{e}} > 80$~GeV and $M_{1/2} > 60$~GeV (this
is a rough bound inferred from the gluino bound $M_{\tilde{g}} \gtrsim
190$~GeV), we find $\mu > 82$~GeV.  This basically does not require any
cancellation, since $M_{Z}^{2}/2\mu^{2} = 65\%$.

The situation can be somewhat ameliorated in the MSSM if there is a
Fayet--Illioupoulos $D$-term for the $U(1)_{Y}$ gauge group.  Such a
$D$-term is known to arise in many ways, such as kinetic mixing
of the $U(1)_{Y}$ and $U(1)_{\it mess}$ gauge fields \cite{D-term}.
The running of
all the parameters remains the same except that one adds another
contribution from the $D_{Y}$ at the weak scale.  If the sign is
appropriate, it increases $m^{2}_{\tilde{e}}\rightarrow
m^{2}_{\tilde{e}} + D_{Y}$ and $m^{2}_{H_{u}}\rightarrow
m^{2}_{H_{u}} + \frac{1}{2} D_Y$
(less negative) while decreasing $m^{2}_{H_{d}}\rightarrow
m^{2}_{H_{d}} - \frac{1}{2} D_{Y}$.  All of these help push
the parameters relevant for electroweak symmetry breaking in the right
direction.  Larger $m^{2}_{\tilde{e}}$ reduces the lower bound on
$\langle F_{S}\rangle/\langle S \rangle$, and a less negative
$m^{2}_{H_{u}}$ is also welcome.  Therefore the sensitivity to $\mu$
(required cancellation between $\mu$ and soft SUSY-breaking parameters) in the
MSSM can be improved in the presence of a $D_{Y}$ with the appropriate
sign.

We will see in the next two sections that the situation in the NMSSM is
much worse.  There is no phenomenologically viable solution
to electroweak symmetry breaking. One can modify the model to
generate a large negative mass-squared for the singlet field and 
then find a viable solution. This solution also requires a cancellation
among parameters which has the same origin as the cancellation
present in the MSSM.  We will also see that the addition of the
Fayet--Illiopoulos $D$-term does not improve the situation within the
NMSSM.

\setcounter {equation}{0}

\section{The NMSSM with the GMSB}
\label{sec:NMSSM}

In this section we study the feasibility of implementing the GMSB in
the framework of the
Next-to-Minimal Supersymmetric Standard Model (NMSSM). We begin
our presentation by introducing the NMSSM: its particle content,
superpotential, and soft SUSY-breaking terms. We briefly review
the major steps in our analysis: the boundary conditions for the
breaking terms, the RG evolution, and the minimization of the
weak-scale one-loop effective potential. We then describe the
results of a numerical scan of a large portion of the model's
parameter space. We find that it is
impossible to evade the present-day experimental constraints. We
further strengthen this argument by providing a semi-analytical
explanation for the inevitability of this conclusion.

\subsection{The NMSSM}

The NMSSM represents an attempt to solve the $\mu$-problem of the
MSSM in the simplest and most direct way: the spectrum of the
MSSM is augmented by a gauge singlet superfield {\it N}, which couples to
$H_d H_u$ and  plays the role of the $\mu$-term once it develops a
nonzero vacuum expectation value \cite{Eetal}. The original $\mu$-term
is banned
from the theory so that there are no dimensionful parameters
left in the superpotential.

The VEV of the scalar component of {\it N} is determined by minimizing the
scalar potential with respect to $H_d$, $H_u$, and $N$ simultaneously. It is
natural to expect the VEVs to be of the same order of
magnitude for all three fields, thus generating an effective
$\mu$-parameter of order the weak scale, as required by
phenomenology.

The complex scalar {\it N} introduces two additional degrees of
freedom to the Higgs sector. Therefore, the particle spectrum of the
NMSSM contains three CP--even Higgs scalars, two CP--odd Higgs
scalars, and one charged Higgs scalar. Immediately, there is a
problem: one of the pseudoscalar Higgs bosons is massless. This happens
because the superpotential $W=\lambda N H_d H_u $ has a
Peccei-Quinn symmetry $N \rightarrow N e^{i \alpha}, H_d H_u
\rightarrow H_d H_u e^{-i  \alpha}$.
This symmetry is spontaneously broken by the VEVs of the fields, making
one of the pseudoscalars massless.

The standard solution to this problem is to introduce a term cubic in
{\it N}, which explicitly breaks the symmetry mentioned above. This
term is allowed by the gauge symmetries of the model and does not
contain a dimensionful coupling constant, so it is generically expected
to be present in the superpotential. One, however, still
has to worry about a light pseudoscalar Higgs boson. As we will show
shortly, its mass can also be small because of the presence of a
different (approximate) $U(1)$ symmetry.

Overall, the only change made to the MSSM superpotential is the
following:
\begin{eqnarray}
\mu H_dH_u \longrightarrow \lambda N H_d H_u - {k\over 3} N^3,
\end{eqnarray}
while the corresponding change to the soft SUSY-breaking part of the
potential is:
\begin{eqnarray}
- m_{3}^{2} (H_{d} H_{u} + c.c.) \longrightarrow  -
(\lambda A_{\lambda} N H_d H_u + \frac{1}{3} k A_k N^3 + h.c. )
+ m_N^2 |N|^2.
\label{nmssmsoftterms}
\end{eqnarray}

One can determine the VEVs of the
Higgs fields $H_d$, $H_u$, and $N$ by minimizing the scalar potential, which
at the tree-level consists of the $F$-terms, $D$-terms, and
soft SUSY-breaking terms:
\begin{eqnarray}
& & V_{\it Higgs}^{\it tree}  =  V_F + V_D + V_{\it soft}, \nonumber \\
& & V_F = |\lambda H_d H_u - k N^2|^2 + \lambda ^2 |N|^2 (|H_d|^2+|H_u|^2),
           \nonumber  \\
& & V_D  =  \frac{{g_2}^2}{8} (H_d^{\dagger} \vec{\sigma}H_d+H_u^{\dagger}
\vec{\sigma}H_u)^2+\frac{g'^2}{8} (|H_d|^2-|H_u|^2)^2, \nonumber \\
& & V_{\it soft} = {m^2_{H_d}} |H_d|^2 + {m^2_{H_u}} |H_u|^2 +{m^2_N}|N|^2 -
 (\lambda A_{\lambda} H_d H_u N + h.c.)-\left(\frac{k}{3} A_{k} N^3 +
h.c.\right).\;\;\;\;\;\;\;\;\;\;\;\;\;
\end{eqnarray}
An important fact to notice is that both $V_F$ and $V_D$ remain
unchanged when $H_d, H_u, \mbox{ and } N$ are all rotated by the same
phase. In fact, only the soft SUSY-breaking
$A$-terms are not invariant under this transformation. This can be
potentially dangerous, because we, in general, consider the $A$-terms
to be zero at the messenger scale, and their sizes at the weak scale are
determined by the RG evolution. If the generated values of
$A_{\lambda}$ and $A_k$ are not large enough, our scalar potential has
an approximate $U(1)$ symmetry. This symmetry is spontaneously broken by
the vacuum expectation values of the Higgs fields, and, as before, we have
to worry about a light pseudoscalar Higgs boson.

We denote the VEVs of the neutral components of the Higgs fields by
$v_d$ and $v_u$, as in Section~\ref{mssmsection}, and the VEV of the
singlet field by {\it x}:
\begin{eqnarray}
\langle N \rangle = x.
\end{eqnarray}
As a function of these VEVs, the potential has the form
\begin{eqnarray}
\label{treepotential}
V_{neutral}^{tree} & =&|\lambda v_d v_u - k x^2|^2 +
\lambda ^2 |x|^2 (|v_d|^2+|v_u|^2) + {m^2_{H_d}} |v_d|^2
+ {m^2_{H_u}} |v_u|^2 +{m^2_N} |x|^2 -
\nonumber \\ && -  (\lambda A_{\lambda} v_d v_u x +
h.c.)-(\frac{k}{3} A_{k} x^3 + h.c.)
 +  \frac{g^2_2+g'^2}{8} (|v_d|^2-|v_u|^2)^2.
\end{eqnarray}

It is well known that some of the Higgs boson masses receive significant
contributions from radiative corrections. In our numerical
analysis we account for that by employing the one-loop effective potential
\begin{equation}
V_{\it neutral}^{1-\it loop} (v_i)= V_{\it neutral}^{\it tree}(v_i, \mu) +
\frac{1}{64\pi^2} \mbox{STr} {\cal M}^4 (v_i)
\left(\log \frac{{\cal M}^2 (v_i)}{\mu^2} - \frac{3}{2} \right).
\label{oneloopeffective}
\end{equation}
In this expression $ {\cal M}^2 (v_i)$ is a field-dependent scalar
mass-squared matrix, and $\mu$ is the $\overline{\rm MS}$
renormalization scale. As we have indicated explicitly, the values
of the various parameters entering $V^{tree}$ depend on the choice of
this scale. To the leading order this dependence is canceled when
the second term on the right hand side of Eq.~(\ref{oneloopeffective}) is
included, and the result of minimizing $V^{1-loop}$ is less
sensitive to the choice of the scale where one stops running the RG equations.
(Canceling out this dependence completely would require calculating
radiative corrections to all orders.)

The matrix $ {\cal M}^2$ depends on the field VEVs $v_i$ through the
Yukawa couplings of the Higgs fields to various other particles. What
plays a crucial role here is not the absolute values of the masses,
but rather the rate of their change as one changes $v_i$.
Therefore, the most important contribution comes
from the field-dependent masses of the top quark and squarks, which
have the largest Yukawa coupling.
Denoting their mass eigenvalues by $m_t$, $m_{\tilde{t}_1}$ and
$m_{\tilde{t}_2}$ respectively, the contribution to $V^{1-loop}$ from
radiative corrections due to these states is
\begin{eqnarray}
\Delta V = \frac{3}{32\pi^2} \left[
m_{\tilde{t}_1}^4 (v_i) \left(
\log \frac{m_{\tilde{t}_1}^2 (v_i)}{\mu^2} - \frac{3}{2} \right)
+ m_{\tilde{t}_2}^4 (v_i) \left(
\log \frac{m_{\tilde{t}_2}^2 (v_i)}{\mu^2} - \frac{3}{2} \right)
 \right. \nonumber \\
\left.- 2 m_t^4 (v_i) \left( \log \frac{m_t^2 (v_i)}{\mu^2} - \frac{3}{2}
\right) \right].
\end{eqnarray}

\subsection{Numerical Analysis}

In models with the GMSB the values of the soft SUSY-breaking terms are
specified at the messenger scale by Eqs.~(\ref{gauginomass}),
(\ref{sfermionmass}) and (\ref{Aterms}). Their values at the weak
scale can be determined by solving the RG equations given in 
\ref{AppendixA}.

The model has five input parameters: $h_t$, $\lambda$, $k$, $B$, and
$n$. (Note that the only dimensionful input parameter is $B$, and its
magnitude will determine the overall scale of the VEVs and the
soft SUSY-breaking masses.) There are, however, two constrains
which must be satisfied at the weak scale: $ v
\equiv\sqrt{v_d^2+v_u^2}=174$ GeV and
$h_t v_u=165 \pm 5$ GeV.\footnote{Notice that this
number is not equal to the top quark pole mass, the
experimentally measured quantity, because of QCD corrections.
The relationship between the two is given, at 1-loop, by
$m_{pole}=\bar{m}(1+{4\over 3}{\alpha_s \over \pi})$.} A common
approach is to use the minimization conditions and RG equations to
solve for the inputs, given a phenomenologically allowed set of
weak--scale outputs. In the case of
a high messenger scale, however, no easily invertible solution for the
RG equations is available. Instead, we simply choose to tackle the problem
numerically. After running down the RG equations and minimizing the Higgs
potential once, we iterate this procedure, each time adjusting the
value of the parameter {\it B} to fix the overall scale of the VEVs and
masses, while simultaneously changing the dimensionless couplings to
correctly reproduce the top quark mass. This iteration process, in fact,
converges fairly quickly.

Using the procedure above, we perform a numerical scan of a large
portion of the parameter space. We study the low-energy particle
spectrum for various messenger scales $\Lambda$, numbers of
messengers {\it n}, and values of the couplings $\lambda
\mbox{ and } k$.
It is interesting to note that it is very easy to generate non-zero
VEVs for $H_d, H_u,$ and $N$, even when $m_N^2$ is a small {\it
positive}\/ number. This is because the terms $|\lambda v_d v_u - k
x^2|^2 \mbox{ and } A_{\lambda} \lambda v_d v_u x$,  when $\lambda
v_d v_u $ and $ k$ are of the same sign, both ``push'' the VEV of
the real component of the singlet away from the origin.
Unfortunately, we find that, for any choice of values of the input
parameters, there are always particles with unacceptably small
masses. To illustrate the situation, we present in Table~\ref{table1}
our numerical results for several representative
points in different ``corners'' of the parameter space. The first
two points represent the typical situation for the case of the 
low-energy GMSB, the next two are representative of the case of the
high-energy GMSB, and the last one explores the extreme case of
$\Lambda=10^{15}$ GeV. Points 1 and 3 have relatively large values
of $k$, while points 2 and 4 have $k\ll 1$. Notice that in the
table we did not consider a similar limiting case for $\lambda$.
This is not a coincidence. It turns out that, for $\lambda
\lesssim 0.2$, the dominant term in the potential is $V_D$, and
$\tan\beta$ is forced to values very close to one. In this case, in
order to correctly reproduce the top quark mass, one is forced to
choose $h_{t}$ at the weak scale such that $h_{t}$ hits the Landau
pole below the GUT scale. We have chosen to list only the cases
where the couplings in the superpotential remain perturbative up to
the GUT scale. We make, however, no such assumption in our analysis
in the next subsection.

\begin{table}
\caption{The numerically determined NMSSM parameters for five sample points in
  the parameter space. Here $m_{h_i}$ and $m_{A_i}$ refer to the
  eigenvalues of the scalar and pseudoscalar Higgs mass matrices
  respectively, and $m_{\tilde{e}}$ denotes the mass of the right-handed
  selectron. The values of $\lambda$, $k$, and $h_{t}$ are given
  at the weak scale. All the other quantities have been defined
  earlier in the text.}
\label{table1}
\bigskip
\noindent{
\begin{tabular}{|c|c|c|c|c|c|c|} \hline
 & \multicolumn{6}{|c|}{Input Parameters} \\ \hline
point &$\Lambda$ (GeV)&$\lambda$&$k$&B (TeV)&$n$&$h_t$  \\
\hline 1 &$5 \times 10^4$   & 0.25 & 0.1 &  6.4 &  1 & 1.12 \\ \hline
2 &$10^{5}$  & 0.28 &$3\times 10^{-4}$& 3.6 &  3 & 1.08 \\ \hline
  3 &$10^{12}$  & 0.32 &   0.3   & 0.99 &  10 & 1.07 \\ \hline
  4 &$10^{12}$  & 0.25 & $3\times10^{-4}$  & 6.0 &  1 & 1.11 \\ \hline
  5 &$10^{15}$  & 0.28 & 0.3 & 6.9 &  1 & 1.07 \\ \hline
\end{tabular}
}
\bigskip

\noindent{
\begin{tabular}{|c|c|c|c|c|c|} \hline
& \multicolumn{5}{|c|}{Soft SUSY-breaking Parameters at the Weak
 Scale} \\ \hline 
 point&$m_{H_u}^2$ (GeV$^2$)&$m_{H_d}^2$ (GeV$^2$)&$m_N^2$ (GeV$^2$) &
$A_{\lambda}$ (GeV)&$A_k$ (GeV)\\ \hline
 1 & $-2.4\times10^3$& $5.3\times10^2$ &
 4.6&$-1.5$&$-4.0\times10^{-3}$\\ \hline 
 2 & $-2.8\times10^3$& $5.7\times10^2$& 6.8
 &$-2.6$&$-6.2\times10^{-3}$ \\ \hline 
  3 & $-3.1\times10^3$& $4.8\times10^2$& 29&$-11.4$&$-0.15$  \\ \hline
  4 & $-2.5\times10^3$& $6.8\times10^2$& 12&$-8.0$&$-0.11$  \\ \hline
  5 & $-2.9\times10^3$&
 $1.0\times10^3$&$-8.1$&$-9.4$&$-6.0\times10^{-3}$ \\ \hline 
\end{tabular}
}
\bigskip

\noindent{
\begin{tabular}{|c|c|c|c|c|c|c|} \hline
& \multicolumn{2}{|c|}{Field VEVs} &
\multicolumn{4}{|c|}{Particle Masses}\\ \hline
point&$\tan\beta$&$x$ (GeV) & $M_3$ (GeV) & $m_{\tilde{e}}$ (GeV)
  & $m_{h_i}$ (GeV)& $m_{A_i}$ (GeV) \\ \hline
1 &1.59&$-3.7$ & 61 &32 & 85, 39, 35 & 51, 1.8\\ \hline
2&1.84&$-3.7$ & 103&35 &87, 48, 38 & 48, 0.2 \\\hline
3&1.97&$-40$ & 94&36 & 87, 53, 28 & 76, 25 \\ \hline
 4&1.63&$-14$ & 57&34 &85, 43, 37 & 44, 0.5 \\ \hline
 5&1.88&$-49$ & 66&40 &88, 50, 27 & 71, 24 \\ \hline
\end{tabular}
}
\end{table}

It can be seen that, in all the cases presented in Table~\ref{table1},
there are particles with unacceptably small masses. The
result for the low-energy GMSB is not new and has been known for several years
\cite{DN}. On the other hand, the situation with a high messenger
scale had not been quantitatively studied in the literature to
date. One expected feature that we indeed see in points 3 and 5 is
the increase of the pseudoscalar Higgs boson mass with $\Lambda$. This
happens because the magnitude of $A_{\lambda}$, generated by running the
RG equations, increases with the messenger scale, and it is $A_{\lambda}$
that breaks the $U(1)$ symmetry of the potential, as discussed
before. Another result that could have been anticipated is the
smallness of the mass of the light pseudoscalar Higgs when $k\ll1$
(points 2 and 4). This is due to the Peccei-Quinn symmetry, which is
restored in this limit. What is surprising is that raising the
messenger scale by 10 orders of magnitude does not bring any other
significant changes to the particle spectrum. The masses of the gluino,
right-handed selectron, and scalar Higgs boson still remain small.

\subsection{Analytical Considerations}

In this subsection we present a rather simple semi-analytical
argument which explains why there can be no phenomenologically
acceptable solution to the NMSSM with the GMSB. We show that if one
assumes that such a solution exists, one arrives at a
contradiction. We also explain some of the features of the
numerical solutions presented in the previous subsection.

Suppose that for some point in the parameter space an acceptable
solution exists. The problem to be addressed is the smallness of
the selectron, gluino, and Higgs masses. We choose to base our
analysis on the right-handed selectron mass constraint. The
magnitude of $m_{\tilde{e}}$ is directly proportional to the size of the
$B$-parameter. In our numerical procedure the value of $B$ is
chosen in such a way that $v=174$ GeV. A typical value of $B$
obtained in this way yields a very small selectron mass
($m_{\tilde{e}}\sim 35$ GeV), gluino mass ($M_3 \lesssim 100$ GeV), and
soft SUSY-breaking masses for the Higgs bosons ($m_{H_u}^2\sim - 3000 $
GeV$^2$, $m_{H_d}^2
\sim 500$ GeV$^2$).

\begin{figure}[tbp]
  \begin{center}
    \leavevmode
    \epsfxsize=0.8\textwidth \epsfbox{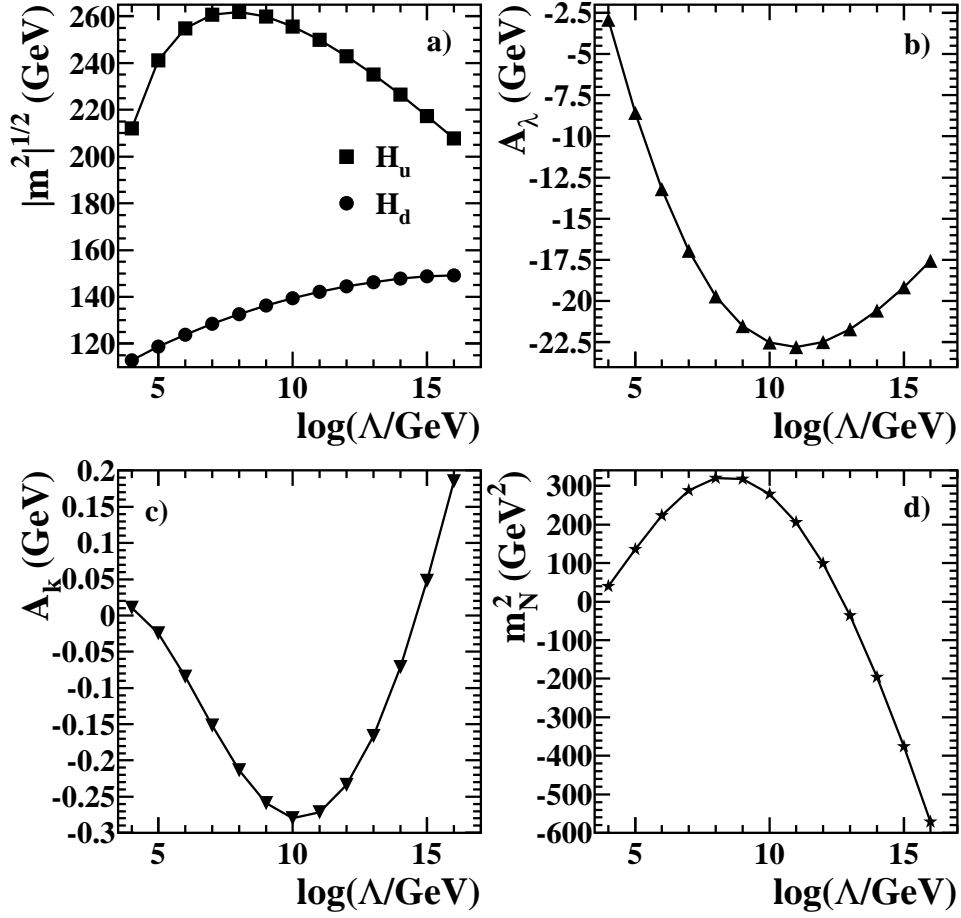}
    \caption[121]{(a) Lower bounds on $|m_{H_u}^2|^{1/2}$ and
      $|m_{H_d}^2|^{1/2}$ as a
      function of the messenger scale $\Lambda$ from
      the selectron mass constraint $m_{\tilde{e}}>80$ GeV. Here $n=1$,
      $h_{t}=1.07$, $k=0.3$ and $\lambda=0.29$ at the weak
      scale. These bounds do not
      change for different values of $k$ or $\lambda$. The other plots
      show typical values of (b) $A_\lambda$, (c) $A_k$, and
      (d) $m_N^2$, for the same choice of parameters that yielded (a).
      The values of these parameters do not change significantly
      for different values of $k$ or $\lambda$.}  
    \label{fig:m2bounds}
  \end{center}
\end{figure}

It is, therefore, obvious that the only chance of obtaining an
acceptable value for the selectron mass is to raise {\it B}, by a
factor of three or more, and try to arrange the other parameters in
such a way that $\sqrt{v_d^2+v_u^2}$ remains 174 GeV. Since {\it B}
feeds into all soft SUSY-breaking masses, their absolute values will
also increase. For example, imposing $m_{\tilde{e}}>80$ GeV forces
$m_{H_u}^2 < -(215 \mbox{ GeV})^2$ for a messenger scale of
$10^{16}$ GeV. For different messenger scales the bound becomes
even more stringent, as shown in Fig~\ref{fig:m2bounds}(a).

To determine the consequences of raising the soft SUSY-breaking masses,
we analyze the Higgs potential (Eq.~(\ref{treepotential})). The
extremization conditions at tree level are
\begin{eqnarray}
   \label{minim_conditions_v1}
\frac{\partial V_{neutral}^{tree}}{\partial v_d} &=& 2(\lambda v_d
v_u-k x^2) \lambda v_u 
+ 2 \lambda^2 x^2 v_d + 2 m_{H_d}^2 v_d - 2 A_{\lambda} \lambda v_u x
+ \frac{g'^2+g_2^2}{4} 2 v_d (v_d^2-v_u^2) , \\
    \label{minim_conditions_v2}
\frac{\partial V_{neutral}^{tree}}{\partial v_u} &=& 2(\lambda v_d
v_u-k x^2) \lambda v_d 
+ 2 \lambda^2 x^2 v_u + 2 m_{H_u}^2 v_u - 2 A_{\lambda} \lambda v_d
x + \frac{g'^2+g_2^2}{4} 2 v_u (v_u^2-v_d^2) , \\
  \label{minim_conditions_x}
\frac{\partial V_{neutral}^{tree}}{\partial x} &=& 2(\lambda v_d v_u-k
x^2)(-2kx) 
+ 2 x \lambda^2 (v_d^2+v_u^2) + 2 m_N^2 x - 2 A_{\lambda} \lambda v_d
v_u - 2 k A_k x^2 .\;\;\;\;\;\;\;\;\;\;\;\;
\end{eqnarray}
The first two equations (Eqs.~(\ref{minim_conditions_v1}) and
(\ref{minim_conditions_v2})) closely resemble the corresponding ones
in the MSSM case. In fact, the only difference in the NMSSM is the
presence of the first term on the right-hand side of
Eq.~(\ref{minim_conditions_v1}) and Eq.~(\ref{minim_conditions_v2}). This
term originates from $|\frac{\partial W}{\partial x}|^2 = |\lambda
v_d v_u-k x^2|^2$ and is, therefore, absent in the MSSM. Dividing
Eq.~(\ref{minim_conditions_v1}) by $v_u$, Eq.~(\ref{minim_conditions_v2})
by $v_d$, and subtracting the two expressions, we can cancel out
this term. As a result, we obtain
\begin{equation}
  \label{mu^2}
  \lambda^2 x^2=- \frac{M_Z^2}{2} -  \frac{1}{2} (m_{H_d}^2+m_{H_u}^2)
   -  \frac{1}{2} \frac {m_{H_d}^2-m_{H_u}^2}{\cos 2 \beta} =
   - \frac{M_Z^2}{2} + \frac{m_{H_d}^2-m_{H_u}^2 \tan^2 \beta}{\tan^2
   \beta-1}\, . 
\end{equation}
Note that this equation is identical to Eq.~(\ref{eq:mueq}) with $\mu
\equiv \lambda x$. To obtain the NMSSM analog of Eq.~(\ref{eq:m3eq}) we divide
Eq.~(\ref{minim_conditions_v1}) by $v_d$,
Eq.~(\ref{minim_conditions_v2}) by $v_u$ and {\it add} the two. Solving
for $A_{\lambda} \lambda x$ ($A_{\lambda} \lambda x \equiv m_3^2$)
we find:
\begin{equation}
  \label{Bmu}
 A_{\lambda} \lambda x = (m_{H_d}^2+m_{H_u}^2+2 \lambda^2 x^2)
 \frac{\sin 2 \beta}{2} + \lambda (\lambda v_d v_u-k x^2)\, .
\end{equation}
Eqs.~(\ref{eq:m3eq}) and (\ref{Bmu}) differ only by the
contribution from $|\frac{\partial W}{\partial x}|^2$.

Eq.~(\ref{mu^2}) states that the value of the effective
$\mu$-parameter generated in this model is subject to a rather
stringent bound: $\lambda^2 x^2 >
-m_{H_u}^2-M_Z^2/2$, which, if one imposes
$m_{H_u}^2 < -(212 \mbox{ GeV})^2$, translates into $\lambda x >
200 \mbox{ GeV}$. Notice that the origin of this bound is the same
as of the bound on the size of the $\mu$-parameter derived in
Section~\ref{mssmsection}, since the condition given by
Eq.~(\ref{mu^2}) is the
same in both cases. In the present case, however, the bound is
stronger because $\tan\beta$ is no longer a free parameter but is
determined by minimizing the Higgs potential.

So far we have only looked at the first two extremization
conditions. We now turn our attention to
Eq.~(\ref{minim_conditions_x}).
Solving for $x^2$ in Eq.~(\ref{mu^2}),
one can rewrite Eq.~(\ref{minim_conditions_x}) as
\begin{equation}
\label{modified_x}
 2{k^2 \over \lambda^2}
\left( \frac{m_{H_d}^2-m_{H_u}^2 \tan^2 \beta}{\tan^2 \beta-1}
-\frac{M_Z^2}{2} \right) = \lambda v^2(k\sin2\beta- \lambda)
- m_N^2 + A_{\lambda} \lambda v^2{\sin2\beta \over {2 x}} + k A_k x\, .
\end{equation}
While we have shown that phenomenology requires the expression in
parenthesis on the left-hand side to be larger than $(200
\mbox{ GeV})^2$, the terms on the right-hand side are all much smaller,
because $m_N^2,\; A_{\lambda}, \mbox{ and } A_k$ are zero at the
messenger scale and the effects of the RG running are relatively
small (see Fig.~\ref{fig:m2bounds}). This means that the above
equation can never be satisfied unless $k\ll\lambda$. 

An immediate consequence of the $ k
\rightarrow 0$ limit is that the mass of the lightest
pseudoscalar Higgs goes to zero, as it becomes a Nambu-Goldstone
boson. (It is for this reason that {\it k} was introduced in the
first place.) Furthermore in the limit of large $\mu$ and
small $k$ the determinant of the scalar Higgs mass-squared matrix becomes
negative, which means that the extremum point given by 
Eqs.~(\ref{minim_conditions_v1}-\ref{minim_conditions_x}) ceases to be a
minimum. To show this we first derive a relationship
between {\it k} and $\sin 2 \beta$. That relationship can be
derived from Eq.~(\ref{mu^2}) and Eq.~(\ref{Bmu}). Neglecting $M_Z^2$,
$A_{\lambda} \lambda x$, and $\lambda^2 v^2$ in comparison to
$m_{H_d}^2$ and $m_{H_u}^2$, we find that
\begin{eqnarray}
  \label{sin2beta}
  \sin 2 \beta = \frac{k}{\lambda} \frac{1}
{1+r^2 \left(\frac{k}{\lambda}\right)^2}
\left(1 + r\sqrt{1+(r^2-1)\left(\frac{k}{\lambda}\right)^2} \right)\, .
\end{eqnarray}
Here $r \equiv
-(m_{H_d}^2+m_{H_u}^2)/(m_{H_d}^2-m_{H_u}^2)$.\footnote{In deriving
  Eq.~(\ref{sin2beta}) it was necessary to 
assume that $\sin 2\beta>k/ \lambda$. This translates into two
requirements: $r>0 \mbox{ and } k/\lambda<1$. We conclude that for
large soft SUSY-breaking Higgs masses-squared it is necessary to have
$\lambda>k$.} For $k\ll \lambda$, Eq.~(\ref{sin2beta}) reduces to
\begin{eqnarray}
  \sin 2 \beta \approx 2 \frac{k}{\lambda}
\left[\frac{-m_{H_u}^2}{m_{H_d}^2-m_{H_u}^2}\right]\, .
\end{eqnarray}

Equipped with the last result, we consider the determinant of the
scalar Higgs mass-squared matrix. The full expression for it is given in
\ref{3x3matrix}; here we only need to identify the leading
terms. We are interested in the case $\mu > \lambda v$, and, as we
have argued before, the soft trilinear couplings $A_{\lambda}$ and
$A_k$ can be neglected. For this reason, the dominant terms will be
the ones containing the highest power of $\mu$:
\begin{equation}
 \label{determ2}
    \det {{\mathcal M }_{scalar}^2} \simeq
    \frac{2 v^2 \mu ^4 }{ \lambda ^3 \sin(2 \beta)}
     \left(- 4 k {\lambda }^4  +  k^3  {{\bar{g}}^2} +
      k^3 \cos (4 \beta ) \bar{g}^2
      + 8 k^2 \lambda ^3 \sin (2 \beta )\right)\, .
\end{equation}
That these are indeed the largest terms was checked numerically.

Taking into account the fact that $ k$ and $\sin 2\beta$ are
proportional to each other for small $k$, one can easily see that,
in the limit $k\rightarrow 0$, the first term dominates and the
determinant is negative.\footnote{Because $\sin 2\beta \propto
  k/\lambda$ there is no ambiguity with sign redefinitions of
  $\lambda$ or $k$ in Eq.~(\ref{determ2}). }

This completes our argument, and we are now able to state that
there can be no phenomenologically viable solution in the context
of the NMSSM. We could have also based our argument on the
gluino mass bound. The experimental constraint $M_3 > 190$ GeV
translates into the requirement $m_{H_u}^2 <-(212\mbox{ GeV})^2$
(assuming $n=1$), and the rest of the argument follows unchanged.
Notice, however, that the bound on $m_{H_u}^2 $ weakens if the
number of messenger fields is taken to be very large.

We now turn to the issue of interpreting the numerical results of
the previous subsection. We would like to understand, for instance,
why the values of the singlet VEV $x$ in Table~\ref{table1} are
always smaller than the VEVs of the Higgs doublets and,
furthermore, why $x$ is only several GeV for a low messenger scale.

The answer comes from considering the extremization condition for
$x$:
\begin{equation}
2{k^2}{x^3} + \lambda \left( \lambda  - k \sin (2\beta ) \right) v^2 x  -
  {\frac{{v^2} \lambda \sin (2 \beta )
      A_{\lambda }}{2}} \simeq 0\, ,
\end{equation}
where we omitted the terms $m_N^2$ and $k A_k x$ ($|m_N^2| \ll
\lambda^2 v^2$ for all the points in the table). For most of the
parameter space the cubic term in $x$ can also be neglected, giving
\begin{eqnarray}
 \label{xA/l}
x \simeq \frac{A_\lambda}{\lambda} {\frac{\sin (2\beta ) }
   {2 \left( 1 -
       {\frac{k}{\lambda }} \sin(2\beta) \right) }}\, .
\end{eqnarray}
Thus the smallness of $x$ is related to the fact that
$A_{\lambda}$ is small. The above approximation holds as long as
\begin{eqnarray}
A_\lambda^2 < \lambda^2 v^2 {\lambda^2 \over k^2} \frac{2\left(1-{k
      \over \lambda}\sin2\beta\right)^3}{\sin^2 2\beta}\, ,
\end{eqnarray}
which is not satisfied only for point 5 in Table~\ref{table1}.
For point 5 the value of $x$ can be approximated by
\begin{eqnarray}
x \simeq \left( A_\lambda v^2\frac{\lambda \sin2\beta}{4 k^2}\right)^{1/3}.
\end{eqnarray}
Again $x<v$ and therefore $\lambda x\ll 175$~GeV. 

Knowing that $x$ is small in this model we can derive another
interesting relation. Neglecting all the terms containing $x$ in
Eq.~(\ref{Bmu}), we obtain:
\begin{eqnarray}
\label{lambdax}
 \lambda^2 v^2 \simeq - (m_{H_d}^2 + m_{H_u}^2)\, .
\end{eqnarray}
This explains why the values of the
soft SUSY-breaking masses for the Higgs bosons are so similar for very
different values of the messenger scale.

Finally, we can say a few words about the scalar Higgs boson
masses. In the limit of small $x$ (and hence small $\mu$), the
dominant term in the determinant of the scalar Higgs mass-squared
matrix (see \ref{AppendixB}) is
\begin{eqnarray}
\det {{\mathcal M }_{scalar}^2} \simeq
{\frac{3 A_{\lambda} {v^6} {{\lambda }^4} {{\bar{g}}^2} }{32 \mu
\sin(2 \beta) }}\, .
\end{eqnarray}
Taking into account the fact that, for small $x$, $\mu\equiv \lambda x
\sim A_\lambda$ (see Eq.~(\ref{xA/l})), the
equation above gives:
\begin{eqnarray}
\label{Higgsestimate}
m_{h_1}^2 m_{h_2}^2 m_{h_3}^2 \sim {\frac{3 {v^6} \lambda^4
{\bar{g}}^2}{32\sin(2 \beta) }}\, .
\end{eqnarray}
This explains why changes in the messenger scale have almost no
effect on the product of the scalar Higgs boson masses
(see Table~\ref{table1}), as long as $\lambda$ is unchanged. For $ \sin(2
\beta)\sim 0.8-0.9$, which is what we typically find in this case,
Eq.~(\ref{Higgsestimate}) gives a ``geometrical average" value of the
scalar Higgs boson mass of only about 50 GeV. This means that, as
long as $x$ is small, the model necessarily yields
phenomenologically unacceptable Higgs boson masses.

\setcounter {equation}{0}

\section{Possible Modifications to the NMSSM}
\label{sec:mod}

In this section we reexamine the expressions derived in
Section~\ref{sec:NMSSM} and attempt to modify the NMSSM to make
it phenomenologically viable. We study several possibilities 
and comment on the problems that arise. Overall, we find none of
these possibilities entirely satisfactory.

\subsection{Extra Vector-like Quarks\label{extraquarks}}
We want to modify the NMSSM in a way that allows one to avoid
the conclusions of Section~\ref{sec:NMSSM}. Recall that the crucial
step in our analysis there was the observation that
Eq.~(\ref{modified_x}) could not be satisfied: the left-hand side was
always greater than the right-hand side. To obtain a consistent
solution one has to somehow make both sides equal. One possibility
is to make $m_N^2$ of the same order of magnitude (and sign) as
$m_{H_u}^2$. That could be accomplished by coupling the singlet to
some new fields and arranging the parameters in such a way that the
SUSY-breaking mass-squared of the singlet is driven sufficiently
negative. This idea was first proposed by Dine and Nelson in Ref.
\cite{DN}, who introduced new color-triplet fields $q'$ and
$\bar{q}'$ and coupled them to {\it N}. The corresponding
superpotential is
\begin{eqnarray}
\label{nmssmmodifiedsuperpotential}
W = h_u Q H_u u^c + h_d Q H_d d^c + h_e L H_d e^c + \lambda
N H_d H_u - \frac{1}{3} k N^3+ \lambda _q N q' \bar{q}'\, .
\end{eqnarray}
According to Eq.~(\ref{sfermionmass}), the scalar components of
$q'$ and $\bar{q}'$ acquire large SUSY-breaking masses, which can
drive $m_N^2$ sufficiently negative.

Agashe and Graesser in Ref.~\cite{AG} did a quantitative study of
this scenario for the case of the low-energy GMSB. They showed that
it is indeed possible to generate a large negative $m_N^2$, in the range  
$- (150 \mbox{ GeV})^2$ to $- (200 \mbox{ GeV})^2$, and further demonstrated
that, with $m_N^2$ of this magnitude, one can choose the input
parameters in such a way that $v=174$ GeV and all
experimental constraints are evaded. They also pointed out that in
this scenario the input parameters need to be fine-tuned in order to
reproduce the above value of $v$. In what follows we give
a set of input parameters that yields an acceptable
particle spectrum, and then proceed to analyze the
sensitivity of the Higgs boson VEVs to the NMSSM coupling constants. 
We clarify the origin of this sensitivity and also extend the
analysis to the case of the high-energy GMSB.  

As an example of an allowed solution, we consider the case of the
low-energy GMSB with $B=50$ TeV, $n=1$, and $\Lambda=100$ TeV. For
$m_N^2=-(190 \mbox{ GeV})^2$, to correctly reproduce $M_Z
\mbox{ and } m_t$ we take $h_t$=0.99, $k=-0.045$ and
$\lambda$=0.11 at the weak scale. We find that $\tan\beta$ equals $-2.9$
for this point. Because the magnitude of the product $Bn$ is now quite
large, the masses of the gluino and right--handed selectron are
safe: $M_3=$ 477 GeV, $m_{\tilde{e}}=$93 GeV. The vacuum expectation
value of the singlet is also large, $x$=2.97 TeV, which, as was
argued earlier, is required by Eq.~(\ref {mu^2}). The eigenvalues of
the scalar Higgs mass matrix are 404, 270, and 90 GeV, and those of
the pseudoscalar Higgs mass matrix are 400 and 6.7 GeV. The last
number appears alarmingly small at first sight but, as shown in
Ref. \cite{AG}, has not been excluded. The reason is that the
corresponding eigenstate $a$ is almost a pure singlet:
\begin{eqnarray}
 |a\rangle=0.031 |H_d \rangle- 0.011 |H_u \rangle-0.999|N \rangle
\end{eqnarray}
The quantitative criterion given in Ref.~\cite{AG}, based on the 
constraint from the $\Upsilon \rightarrow a\gamma$ decay, is  
\begin{eqnarray}
  \frac{\sin 2 \beta \tan \beta}
  {\sqrt{\left( \frac{x}{250~{\rm GeV}} \right)^2+ \sin^2 2 \beta }}
  < 0.43,
\end{eqnarray}
and for the parameter set above the left-hand side equals 0.15.

\begin{figure}[tbp]
  \begin{center}
    \leavevmode
    \epsfxsize=0.6\textwidth \epsfbox{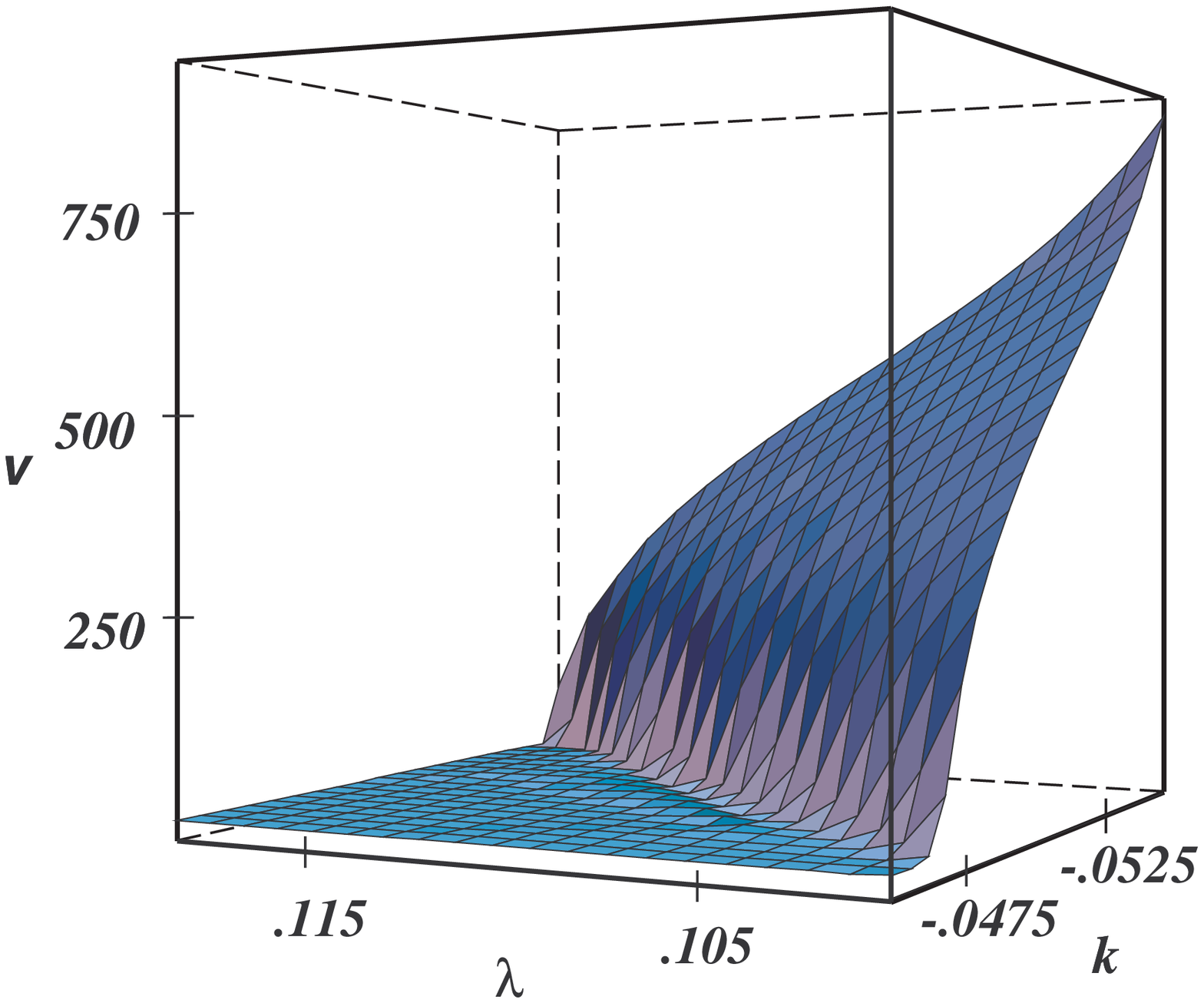}
    \caption[121]{The value of $v\equiv \sqrt{v_d^2+v_u^2}$
      as a function of $\lambda$ and $k$.  The inputs are $n=1$,
      $m_N^2=-(190~\mbox{GeV})^2$, $B=50$~TeV, $\Lambda=100$~TeV,
      $h_t=0.99$.}
    \label{fig:3dplot}
  \end{center}
  \begin{center}
    \leavevmode
    \epsfxsize=0.6\textwidth \epsfbox{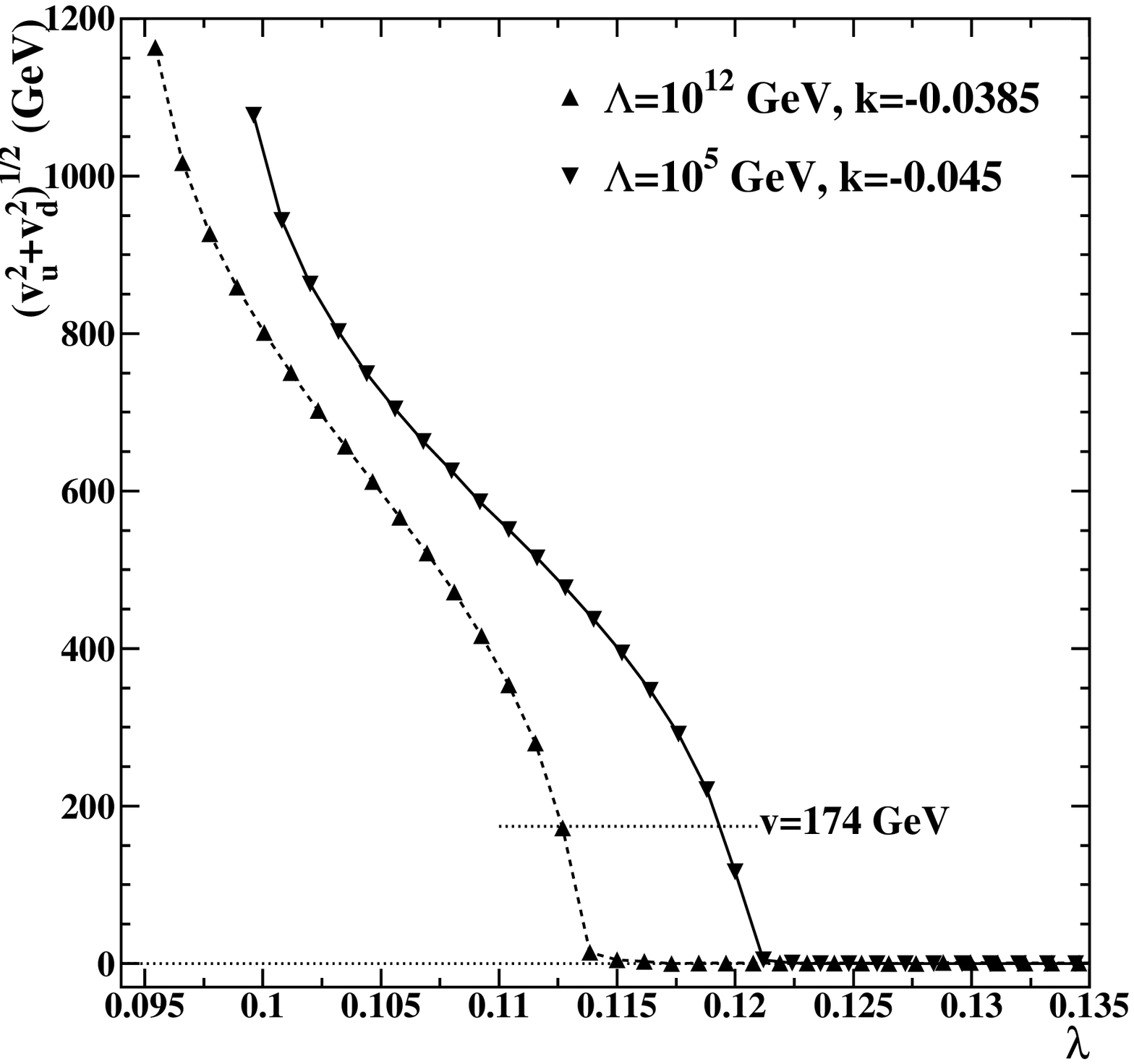}
    \caption[121]{The dependence of $v$
      on the value of $\lambda$ for the high- and low-energy GMSB.  The
      other input parameters are the same as in
      Fig.~\ref{fig:3dplot}.}
    \label{fig:2plots}
  \end{center}
\end{figure}

In this scheme it is, therefore, possible to find a point in the
parameter space which leads to a phenomenologically
viable solution. Unfortunately, as we already mentioned, this solution
is very sensitive to the choice of the superpotential coupling
constants $\lambda$ and $k$. In the remainder of this subsection we
discuss this issue in detail.

The values of the parameters for the set that we have just described
had to be chosen in such a way that the top quark and
$Z$--boson masses were fixed at their known experimental values. It is
interesting to investigate what values of $M_Z$ would be predicted for a
generic choice of the parameters. In Fig.~\ref{fig:3dplot} we plot the
magnitude of the quantity $v\equiv \sqrt{v_d^2+v_u^2}$ as a function
of $\lambda$ and $k$. The figure shows that small changes in both
$\lambda$ and $k$ lead to large changes in $v$. This is
very similar to the situation in the MSSM which was considered in
Section~\ref{mssmsection}. There we showed that the value of the
$\mu$-parameter had to be chosen very carefully in order to yield
the correct value of $v$. In the present case, the points in the
parameter space that correspond to values of $v$ around 174 GeV
lie in  a very thin band on the $\lambda-k$ plane. Also notice that, for this
range of $\lambda$ and $k$, the slope is the steepest. (See 
\ref{commentonprob} for comments on this point.)

It is possible to perform the same type of analysis for a higher
messenger scale. The same problem is found in that case as well. In
Fig.~\ref{fig:2plots} we plot the dependence of $v$ on $\lambda $ for fixed
values of $k$. For comparison, the curve for $\Lambda=
10^{12}$ GeV is plotted next to the curve for $\Lambda= 100$ TeV.
{}From the slopes of these curves one can determine the degree of
sensitivity with respect to $\lambda$, using the definition in
Section~\ref{mssmsection}. The degree of sensitivity, given by ${\rm d}(\log
v)/ {\rm d}(\log\lambda)$, is $2\%$ for the
low-energy curve and $1\%$ for the high-energy curve.  Our numerical
results agree with those in Ref.~\cite{AG} for the low-energy GMSB if the
same inputs parameters are used.

In order to understand this behavior, we once again turn to the
extremization conditions
Eqs.~(\ref{minim_conditions_v1}-\ref{minim_conditions_x}). 
First, we present some qualitative observations. Recall that
phenomenology requires $|x|$ to be rather large
(of the order $\sqrt{|m_{H_u}^2|}/\lambda\gtrsim 1$~TeV), while $v$ 
has to remain
``small'' ($v=174$~GeV) to correctly reproduce $M_{Z}$. As a result, the
terms containing high powers of $x$ and the terms containing
$m_i^2$ ($i={H_d,H_u,N}$) dominate, while the terms with $v_u$ and
$v_d$ are not fixed, and have to absorb the residual
difference between the dominant terms. Therefore, small percentile 
changes in the dominant
terms can result in large percentile changes in the Higgs boson VEVs. 
This is to be contrasted
with the situation in the previous section, where $\lambda^2v^2$
was tied to the value of the sum $m_{H_d}^2+m_{H_u}^2$ (see
Eq.~(\ref{lambdax})).

Next, we try to identify the main source of
this sensitivity. We first consider the dependence of $v$ on
$\lambda$ for fixed $B$, $k$, and $h_t$.
One can use Eq.~(\ref{mu^2}) to solve for $v^2$ and
then isolate the largest contribution to $\partial v/\partial\lambda$.
\begin{eqnarray}
\label{variation}
\frac{\partial v^2}{\partial \lambda}={4 \over \bar{g}^2}\left[-2\lambda x^2
 -2\lambda^2x{\partial x \over \partial \lambda}+
{\partial \over \partial \tan\beta}
 \left( \frac{m_{H_d}^2-m_{H_u}^2 \tan^2\beta}{\tan^2\beta-1}\right)
 {\partial \tan\beta \over \partial \lambda}\right.\nonumber\\
 \left.+\frac{1}{\tan^2\beta-1}{\partial m_{H_d}^2\over \partial\lambda}
 -\frac{\tan^2\beta}{\tan^2\beta-1}{\partial m_{H_d}^2\over
   \partial\lambda}
\right] \, .
\end{eqnarray}
Using the data that led to Fig.~\ref{fig:3dplot}, we numerically
evaluate the derivative around the point $\lambda=0.11,\; k=-0.045$. The
following are the results of evaluating each of the terms on the
right-hand side, respectively:
$-1.4\times10^7,\;-2.4\times10^6,\;-3.9\times10^6,\;
-1.2\times10^3,\;1.9\times10^1$ (GeV$^2$).
The largest term is the first one, the next two
terms combined provide a 45\% correction, and the derivatives of
the soft SUSY-breaking masses can be completely neglected.
In \ref{estimatingsensitivity} we show how these numbers can be
understood by studying the minimization conditions. 

The fact that the dominant contribution to $\partial
v/\partial\lambda$ comes from the first term in Eq.~(\ref{variation})
has a very important implication. It means that the problems of
cancellation in the NMSSM and the MSSM are not merely similar, but
have {\it exactly}\/ the same origin. Indeed, Eq.~(\ref{mu^2})
is the same as Eq.~(\ref{eq:mueq}), and, because in the NMSSM 
$ v$ depends on $\lambda$ mainly through the combination
$\lambda x$, which plays the role of the $\mu$--term, the two models
require roughly the same degree of cancellation.  The degree of
cancellation quoted in Section~\ref{mssmsection} for the MSSM is most
conservatively 16\%, but this is so because one can choose $\tan\beta$
freely in the MSSM.  On the other hand, $\tan\beta$ is determined by
minimizing the potential for the NMSSM and cannot be chosen
arbitrarily to ease the cancellation.  For the value of
$\tan\beta$ which we obtained in the NMSSM, the degree of
cancellation is actually comparable (order a few percent) in the MSSM.
The small  
difference between the two models is due to the 
dependence of $x$ and $\tan\beta$ on $\lambda$.

We have discussed the $\lambda$ dependence of the Higgs boson VEV, and
now turn to the $k$ dependence. Fig. \ref{fig:3dplot} shows that the
points that yield $v=174$ GeV form an almost straight line
on the $\lambda-k$ plane. It can be shown (see
\ref{estimatingsensitivity}) that in order to
keep $v$ constant one has to change $k$ and $\lambda$ according to
$\Delta k/k=\Delta\lambda/\lambda$. The sensitivity of $v$ to $k$ is,
thus, related to the sensitivity of $v$ to $\lambda$, which, in turn,
originates from the need to carefully choose the $\mu$-parameter in
the MSSM as discussed in Section \ref{mssmsection}.

To summarize, we have shown that this model requires a very
particular choice of parameters to yield the correct $Z$-boson
mass. Furthermore, we explained that the sensitivity of the $Z$-boson
mass to the NMSSM couplings has the same origin as the sensitivity of
the $Z$-boson mass to the value of the $\mu$-parameter in the MSSM. We
emphasize that the problem is present for both high and low messenger
scales, simply because the bound on the $\mu$-parameter does not
weaken as one raises the messenger scale.

\subsection{Hypercharge $D$-term}

Next, we investigate what happens if the $D$-term contributions
described at the end of Section~\ref{mssmsection} are included. First, we
consider the case of the NMSSM with no extra particles added. We try
to determine if, by introducing the $D$-terms, it is possible to make
$v$ smaller. If that happened, {\it v} could be rescaled back by
increasing $B$, and that would raise all masses in the model, as
desired. We find that this is not the case. Upon adding the $D$--terms
both $\tan \beta$ and $\langle x\rangle$ change, but $v_d^2+v_u^2$,
curiously enough, remains virtually constant. This happens because, in
the limit $x^2 \ll v^2$, $v^2$ is constrained by Eq.~(\ref{lambdax}),
and the change
$m^{2}_{H_{d}} \rightarrow m^{2}_{H_{d}} - \frac{1}{2} D_{Y}$,
$m^{2}_{H_{u}} \rightarrow m^{2}_{H_{u}} + \frac{1}{2} D_{Y}$
preserves the quantity  $m^{2}_{H_{d}} + m^{2}_{H_{u}}$.

The next question to ask is whether the $D$-terms can decrease the degree of
cancellation for the case with $q'$ and $\bar{q}'$ added.
The answer is again negative and the reason can be seen from
Eq.~(\ref{Bmu}). Recall that the degree of cancellation is controlled by
the magnitude of $x^2$. As long as $A_\lambda\lambda x$ and 
$\lambda^2 v_d v_u$ can be neglected compared to
$m_{H_d}^2+m_{H_u}^2$, Eq.~(\ref{Bmu}) yields
\begin{eqnarray}
x^2 \simeq -\frac{(m_{H_d}^2+m_{H_u}^2)}{2\lambda(\lambda-{k
    \over \sin2\beta})} \, ,
\end{eqnarray}
and the relevant quantity is again
$m^{2}_{H_{d}} + m^{2}_{H_{u}}$.

\subsection{Large Trilinear Couplings}

At last, we consider the scenario proposed by Ciafaloni and Pomarol
\cite{CP}. They consider a modified version of the NMSSM, where $k=0$,
$\lambda\ll 1$ and the value of $A_{\lambda}$ is large at the
messenger scale. Their model also contains, in the 
potential at the weak-scale, a 
linear term in $N$ which is generated by tadpole diagrams and solves the
problem of a light pseudoscalar. They find that the requirement of the
positivity of the 
determinant of the scalar Higgs boson mass-squared matrix is very
restrictive. We repeat part of their analysis to determine if their
choice of parameters could indeed lead to a phenomenologically viable
electroweak symmetry breaking spectrum. Note that, as far as the
following is concerned, their model is identical to the NMSSM.

The full expression for the determinant can be found in 
\ref{3x3matrix}. In the limit of $k\rightarrow 0$ and
$\lambda \rightarrow 0$
\begin{equation}
\det {{\mathcal M }_{scalar}^2} \simeq 
\frac{A_{\lambda}^2 v^2 M_Z^2\lambda^2 \sin^2(4\beta) }{4(1+y)^3}
\left[1+y-{y^2 \over \cos^2(2\beta)}\left({A_{\lambda}^2 \over
      M_Z^2}+1\right) - {y^3 \over \cos^2(2\beta)}\right]\, ,
\end{equation}
where we introduced a variable $y \equiv \bar{g}^2 m_N^2/ (2 M_W^2)$
to conform to the notation used in Ref. \cite{CP}. From the extremization
conditions for the potential,
Eqs.~(\ref{minim_conditions_v1}-\ref{minim_conditions_x}), one can
show that 
$\mu=A_{\lambda} \sin(2\beta)/(2(1+y))$.
There are two intervals of $y$ over which the determinant is greater
than zero. One interval is where both the expression in the brackets
and the denominator are positive. It is given approximately by
the following bound on $|y|$:
\begin{equation}
|y|<\left|\cos 2\beta\left(1+{A_{\lambda}^2 \over
      M^2_Z}\right)^{-1/2}\right|\, . 
\end{equation}
The other interval, not mentioned in \cite{CP}, 
is approximately $\left(-({A_{\lambda}^2 \over  M_Z^2}+1), -1\right)$, where
 both the denominator and the bracketed expression are negative.

The first interval, for $A_{\lambda} > M_Z$, corresponds to rather small
values of $m_N^2$ and 
\begin{equation}
 \mu \simeq {1 \over 2} A_{\lambda} \sin(2\beta)\, .
\end{equation}
Using this equation together with Eq.~(\ref{Bmu}), one can derive the
following result:
\begin{equation}
\label{pomeq}
  A_{\lambda}^2 \left(1-{\sin^2(2\beta) \over 2} \right)\simeq
   (m_{H_u}^2+m_{H_d}^2)\, .
\end{equation}
The above equation is impossible to satisfy in models with the GMSB,
because the combination $(m_{H_u}^2+m_{H_d}^2)$ is always
negative at the weak scale for the messenger-scale
boundary conditions given by Eqs.~(\ref{gauginomass}) and
(\ref{sfermionmass}). To satisfy Eq.~(\ref{pomeq}), a drastic modification
of the boundary conditions would be required.

We now turn our attention to the second possibility. It requires a
relatively large negative value of the singlet soft SUSY-breaking mass-squared:
$ m_N^2 < - 2/(\bar{g}^2) \times M_W^2 = - (132 \mbox{ GeV})^2$. This
value is impossible to generate unless, as before, one introduces fields $q'$
and $\bar{q}'$ and couples them to $N$. Even with the introduction of
these fields, if $k=0,\;\lambda\ll 1 $, the extremization
conditions cannot be simultaneously satisfied. This can be seen in the
following way. For $k=0$ Eq.~(\ref{modified_x}) takes on the form
\begin{equation}
 x= \frac{A_{\lambda} \lambda v^2 \sin(2\beta)}{m_N^2+\lambda^2 v^2} \, ,
\end{equation}
which implies $x\rightarrow 0$ as $\lambda \rightarrow 0$. This is
incompatible with Eq.~(\ref{mu^2}), which requires that $x \rightarrow
\infty$ as 
$\lambda \rightarrow 0$.

\section{Conclusion}

We studied the issue of electroweak symmetry breaking in models with
the gauge mediation of supersymmetry breaking (GMSB).  We first reviewed
various proposals in the literature to generate the $\mu$-parameter of
the MSSM with the same order of magnitude as the soft SUSY-breaking
parameters such 
as squark, slepton, and gaugino masses.  We find that most of them require
small parameters which are accidentally of the same magnitude as the
loop factors, cancellation of the kinetic mixing terms at the
level of $10^{-4}$, omission of interactions allowed by symmetries, or
many new degrees of freedom not motivated otherwise.

Even if one could generate the $\mu$-parameter with the same order of magnitude
as the soft SUSY-breaking parameters, it has to have particular values
to reproduce $M_Z = 91$~GeV.  We studied this question numerically and
found the following.  The current experimental lower bounds on
superparticle masses limit the overall scale of SUSY breaking from
below, which in turn limits $m_{H_{u}}^{2}< 0$ from above ({\it
i.e.}\/, $|m_{H_{u}}^{2}|$ from below).  To reproduce $M_{Z}$,
$\mu^{2}$ needs to cancel (too-negative) $m_{H_{u}}^{2}$ and is hence bounded
from below.  Therefore, there is some cancellation required between
$\mu^{2}$ and $m_{H_{u}}^{2}$.  Even with the most
conservative set of parameters, we found that a cancellation of $16$\% is
necessary. The situation is worse for most of the parameter
space.  This situation was contrasted to the supergravity scenario
where the current experimental lower bounds on superparticle masses do
not require a significant cancellation among parameters.

The simplest mechanism to generate the $\mu$-parameter would be the NMSSM,
the minimal extension of the MSSM without
dimensionful parameters in the superpotential.  The NMSSM is known not
to work with the low-energy GMSB, but there was hope that it
may work with higher messenger scales.  We have shown that this is
unfortunately not the case.  The current bounds on the superparticles
masses are already strong enough to exclude the model completely.  We
presented a semi-analytic discussion to clarify why the NMSSM fails.

We also discussed various possible modifications to the NMSSM and
whether they could lead to a viable electroweak symmetry breaking.  The
introduction of extra vector-like quarks coupled to the
NMSSM singlet produces a large negative mass-squared for the singlet,
and leads to a viable electroweak symmetry breaking.  One needs to
adjust the parameters to a few percent, which is comparable to the
MSSM case for the same $\tan\beta$ range.
A Fayet--Illiopoulos $D$-term for $U(1)_{Y}$ does not improve the
situation.

The overall prospect of electroweak symmetry breaking with the GMSB
remains unclear.  We hope our detailed investigation
prompts further studies on this issue.

\section*{Acknowledgments}

We would like to thank Kaustubh Agashe and Michael Graesser for useful
discussions. We would also like to thank Gian Giudice and 
Riccardo Rattazzi for helping
us understand the mechanism for generating a $\mu$-term proposed in
references \cite{DGP} and \cite{DDR}. This
work was supported in part by the Director, Office of Energy
Research, Office of High Energy and Nuclear Physics, Division of
High Energy Physics of the U.S.  Department of Energy under Contract
DE-AC03-76SF00098 and in part by the National Science Foundation
under grant PHY-90-21139.  AdG was also supported by CNPq (Brazil).
HM was also supported by the Alfred P. Sloan Foundation.

\appendix

\renewcommand{\thesection}{Appendix \Alph{section}}

\section{The Renormalization Group Equations of the NMSSM}
\label{AppendixA}

\renewcommand{\theequation}{\Alph{section}.\arabic{equation}}

In this appendix we list all of the RG equations for the NMSSM,
at 1-loop \cite{NMSSMreg}. These are the equations used, in section 4,
to determine the coupling constants and SUSY-breaking parameters of
the NMSSM at the weak scale, given their values at the messenger
scale.

\begin{eqnarray}
16\pi^2\frac{d}{dt} g'     & = & 11g'^3 , \\
16\pi^2\frac{d}{dt} g_2     & = & g_2^3 , \\
16\pi^2\frac{d}{dt} g_3     & = & (-3)g_3^3  ,\\
16\pi^2\frac{d}{dt} h_t     & = & (6 h_t^2 + h_b^2 + \lambda^2
                   - \frac{13}{9} g'^2
                   - 3 g_2^2 - \frac{16}{3} g_3^2) h_t ,\\
16\pi^2\frac{d}{dt} h_b     & = & (6 h_b^2 + h_t^2 + h_\tau^2 + \lambda^2
                   - \frac{7}{9} g'^2 - 3 g_2^2
                   - \frac{16}{3} g_3^2) h_b ,\\
16\pi^2\frac{d}{dt} h_\tau  & = & (4 h_\tau^2 + 3 h_b^2 + \lambda^2
                   - 3 g'^2 - 3 g_2^2) h_\tau ,\\
16\pi^2\frac{d}{dt} \lambda & = & (4 \lambda^2 + 2 k^2 + 3 h_t^2
                   + 3 h_b^2 + h_\tau^2
                   - g'^2 - 3 g_2^2) \lambda ,\\
16\pi^2\frac{d}{dt} k       & = & 6 (\lambda^2 + k^2) k .
\end{eqnarray}
In the above equations $g'$ is the $U(1)_{Y}$ gauge coupling;
explicitly $g'=e /\cos \theta_W$. $g_{2}$ and $g_{3}$ are,
respectively, the weak and strong coupling constants.
One defines $g_1$ to be the hypercharge coupling constant in the
GUT normalization, {\it i.e.}\/ $g_1 \equiv \sqrt {\frac{5}{3}} g'$ and
$\alpha_1 \equiv \frac{5}{3} \alpha'$.
Gauge couplings at the messenger scale are defined in such a way that
they match their experimental values at the $Z$-mass.
We only consider the effect of third generation Yukawa couplings,
namely, $h_{t}$, $h_{b}$ and $h_{\tau}$.

\begin{eqnarray}
16\pi^2\frac{d}{dt} A_{u_a}  & = & 6 h_t^2 (1+\delta_{a 3}) A_t
                 + 2 h_b^2 \delta_{a 3} A_b
                  + 2 \lambda^2 A_\lambda \nonumber \\
            & - & 4 (\frac{13}{18} g'^2 M_1 + \frac{3}{2} g_2^2 M_2
                       + \frac{8}{3} g_3^2 M_3) ,\\
16\pi^2\frac{d}{dt} A_{d_a} & = & 6 h_b^2 (1+\delta_{a 3}) A_b
                  + 2 h_t^2\delta_{a 3} A_t
                  + 2 h_\tau^2 \delta_{a 3} A_\tau
                  + 2 \lambda^2 A_\lambda \nonumber \\
            & - & 4 (\frac{7}{18} g'^2 M_1 + \frac{3}{2} g_2^2 M_2
                       + \frac{8}{3} g_3^2 M_3) ,\\
16\pi^2\frac{d}{dt} A_{e_a} & = & 2 h_\tau^2 (1 + 3 \delta_{a 3}) A_\tau
                +  6 h_b^2 A_b + 2 \lambda^2 A_\lambda \nonumber \\
            & - & 6 (g'^2 M_1 + g_2^2 M_2) ,\\
16\pi^2\frac{d}{dt} A_\lambda & = & 8 \lambda^2 A_\lambda - 4 k^2 A_k
                    + 6 h_t^2 A_t
                    + 6 h_b^2 A_b + 2 h_\tau^2 A_\tau \nonumber \\
            & - & 2 (g'^2 M_1 + 3 g_2^2 M_2) ,\\
16\pi^2\frac{d}{dt} A_k     & = & 12 (k^2 A_k - \lambda^2 A_\lambda).
\end{eqnarray}
$A_{i}$ are the soft SUSY-breaking trilinear couplings, given in Sections~2
and 4. Note that we only consider third generation trilinear
couplings, namely $A_t h_t= {\cal A}_{u}^{33}$, $A_b h_b= {\cal
  A}_{d}^{33}$, $A_\tau h_\tau= {\cal A}_{l}^{33}$.  
$M_i$ ($i$=1,2,3) are the soft SUSY-breaking gaugino masses and they
evolve, at one 
loop, identically to $\alpha_i$. Explicitly
\begin{eqnarray}
\frac{M_i(Q)}{M_{\frac{1}{2}}} = \frac{g_i^2(Q)}{g_X^2},\
\end{eqnarray}
where $g_{X}$ is the value of all $g_{i}$ at the GUT scale, while
$M_{\frac{1}{2}}$ is the common gaugino mass at the GUT scale.
\begin{eqnarray}
16\pi^2\frac{d}{dt} m_{\tilde{Q}_a}^2 & = & 2 \delta_{a 3} h_t^2
                    (m_{\tilde{Q}_3}^2 + m_{H_u}^2 + m_{\tilde{t}}^2 + A_t^2)
                  + 2 \delta_{a 3} h_b^2
                    (m_{\tilde{Q}_3}^2 + m_{H_d}^2 + m_{\tilde{b}}^2 +
                    A_b^2)
                    \nonumber \\
              & - & 8 (\frac{1}{36} g'^2 M_1^2 + \frac{3}{4} g_2^2 M_2^2
                      +\frac{4}{3} g_3^2 M_3^2)
                    + \frac{1}{3} g'^2 \xi ,\\
16\pi^2\frac{d}{dt} m_{\tilde{u}_a}^2 & = & 4 \delta_{a 3} h_t^2
                    (m_{\tilde{Q}_3}^2 + m_{H_u}^2 + m_{\tilde{t}}^2 +
                    A_t^2)
                    \nonumber \\
              & - & 8 (\frac{4}{9} g'^2 M_1^2 + \frac{4}{3} g_3^2 M_3^2)
                    - \frac{4}{3} g'^2 \xi, \\
16\pi^2\frac{d}{dt} m_{\tilde{d}_a}^2 & = & 4 \delta_{a 3} h_b^2
                    (m_{\tilde{Q}_3}^2 + m_{H_d}^2 
                    + m_{\tilde{b}}^2 + A_b^2) \nonumber \\
              & - & 8 (\frac{1}{9} g'^2 M_1^2 + \frac{4}{3} g_3^2 M_3^2)
                    + \frac{2}{3} g'^2 \xi ,\\
16\pi^2\frac{d}{dt} m_{\tilde{L}_a}^2 & = & 2 \delta_{a 3} h_\tau^2
                    (m_{\tilde{L}_3}^2 + m_{H_d}^2 
                    + m_{\tilde{\tau}}^2 + A_\tau^2) \nonumber \\
              & - & 8 (\frac{1}{4} g'^2 M_1^2 + \frac{3}{4} g_2^2 M_2^2)
                    - g'^2 \xi ,\\
16\pi^2\frac{d}{dt} m_{\tilde{e}_a}^2 & = & 4 \delta_{a 3} h_\tau^2
                    (m_{\tilde{L}_3}^2 + m_{H_d}^2 
                    + m_{\tilde{\tau}}^2 + A_\tau^2) \nonumber \\
              & - & 8 g'^2 M_1^2 + 2 g'^2 \xi ,\\
16\pi^2\frac{d}{dt} m_{H_d}^2 & = & 6 h_b^2
                    (m_{\tilde{Q}_3}^2 + m_{H_d}^2 + m_{\tilde{b}}^2 + A_b^2)
                  + 2 h_\tau^2
                    (m_{\tilde{L}_3}^2 + m_{H_d}^2 
                    + m_{\tilde{\tau}}^2 + A_\tau^2) \nonumber \\
              & + & 2 \lambda^2
                    (m_{H_d}^2 + m_{H_u}^2 + m_N^2 + A_\lambda^2)
                   - 8(\frac{1}{4} g'^2 M_1^2 + \frac{3}{4} g_2^2 M_2^2)
                   - g'^2 \xi ,\\
16\pi^2\frac{d}{dt} m_{H_u}^2 & = & 6 h_t^2
                   (m_{\tilde{Q}_3}^2 + m_{H_u}^2 + m_{\tilde{t}}^2 + A_t^2)
                  + 2 \lambda^2
                   (m_{H_d}^2 + m_{H_u}^2 + m_N^2 + A_\lambda^2)
                    \nonumber \\
              & - & 8 (\frac{1}{4} g'^2 M_1^2 + \frac{3}{4} g_2^2 M_2^2)
                  + g'^2 \xi ,\\
16\pi^2\frac{d}{dt} m_{N}^2   & = & 4 \lambda^2
                    (m_{H_d}^2 + m_{H_u}^2 + m_N^2 + A_\lambda^2)
                   + 4 k^2 (3 m_N^2 + A_k^2).
\end{eqnarray}
$\xi$ is the
hypercharge-weighted sum of all soft SUSY-breaking masses-squared
\begin{eqnarray}
\xi = \sum_i Y_i m_i^2,
\end{eqnarray}
where $i$ runs over all scalar particles.
With the boundary conditions in Eqs.~(\ref{gauginomass},\ref{sfermionmass}),
$\xi=0$ and remains zero throughout the RG evolution.
All soft SUSY-breaking mass-squared terms were
taken to be diagonal.  Again, we only consider the running of third
generation soft SUSY-breaking masses-squared. $m_{N}^{2}$ is defined
in Section~\ref{sec:NMSSM}. 

\setcounter{equation}{0}
\section{Scalar Higgs Mass-Squared Matrix}
\label{AppendixB}

In this appendix we explicitly show the $3\times 3$ scalar Higgs
mass-squared matrix of the NMSSM.
\label{3x3matrix}
\begin{eqnarray}
\lefteqn{
{{\mathcal M }_{scalar}^2} ={1 \over 2} \frac{\partial^2
V_{tree}}{\partial v_i
\partial v_j}= {1 \over 2} \times} \nonumber \\
 & & \!\!\!\! \!\!\!\!\left(\begin{array}{ccc}
  {{\bar{g}}^2}{v_1^2} + \left(A_{\lambda}+\frac{k \mu }{\lambda}\right)
  \frac{2 \mu v_2}{v_1} &
  (4 \lambda ^2 - \bar{g}^2) v_1 v_2 - 2 \mu \left(A_{\lambda}+
     \frac{ k \mu }{\lambda } \right)&
   4 \lambda  \mu  {v_1} - 2 A_{\lambda} \lambda  {v_2} - 4 k \mu  {v_2} \\
    (4 \lambda ^2 - \bar{g}^2) v_1 v_2 - 2 \mu \left(A_{\lambda}+
    \frac{ k \mu }{\lambda } \right)&
   \bar{g}^2 v_2^2 +
    \left(A_{\lambda}+ \frac{k \mu}{\lambda} \right) \frac{2  \mu  v_1}{v_2}&
   -2 A_{\lambda} \lambda  {v_1} - 4 k \mu  {v_1} +
    4 \lambda  \mu  {v_2} \\
   4 \lambda  \mu  {v_1} - 2 A_{\lambda} \lambda  {v_2} -
    4 k \mu  {v_2}&
   -2 A_{\lambda} \lambda  {v_1} - 4 k \mu  {v_1} +
    4 \lambda  \mu  {v_2}&{\frac{-2 A_k k
        \mu }{\lambda }} + {\frac{8 {{k}^2} {{\mu }^2}}
      {{{\lambda }^2}}} + {\frac{2 A_{\lambda} {{\lambda }^2} {v_1} {v_2}}
      {\mu }}
 \end{array}
 \right), \nonumber\\
\end{eqnarray}
where $v_{i}$ for $i=1,2,3$ corresponds, respectively, to $v_{d}$,
$v_{u}$ and $x$. All other parameters were defined in previous sections.

The determinant of the matrix above can be evaluated explicitly,
and its full expression is given bellow. Various limits of this
determinant are considered in the body of the paper.
\begin{eqnarray}\label{fulldeterminant}
& &\det {{\mathcal M }_{scalar}^2} =\nonumber \\
  & & {\frac{{v^2} }{32 {{\lambda }^3} \mu \sin(2 \beta) }}
     \left( -6 A_{\lambda} v^4 {\lambda }^9 - 32 A_{\lambda}^3
  {\lambda }^5 {\mu }^2 - 
       64 A_{\lambda} v^2 {\lambda }^7 {\mu }^2 - 32 A_k A_{\lambda} k
        {\lambda }^4 {\mu }^3 - 160 A_{\lambda}^2 k {\lambda }^4 {\mu }^3 -
        \right. \nonumber \\
       & & 128 k v^2 {\lambda }^6 {\mu }^3 - 32 A_k k^2 {{\lambda }^3}
  {{\mu }^4} - 
       128 A_{\lambda} k^2 {{\lambda }^3}
        {{\mu }^4} - 256 A_{\lambda} {{\lambda }^5} {{\mu }^4} -
       256 k {\lambda }^4 {{\mu }^5} +
       8 A_{\lambda} {v^4} {{\lambda }^9} \cos (4 \beta ) + \nonumber \\
     & &  32 {A_{\lambda}^3} {{\lambda }^5} {{\mu }^2}
        \cos (4 \beta ) +
       64 A_{\lambda} {v^2} {{\lambda }^7} {{\mu }^2}
        \cos (4 \beta ) +
       32 A_k A_{\lambda} k
        {{\lambda }^4} {{\mu }^3} \cos (4 \beta ) +
       160 {A_{\lambda}^2} k {{\lambda }^4}
        {{\mu }^3} \cos (4 \beta ) + \nonumber \\
      & & 128 k v^2 {{\lambda }^6}
        {{\mu }^3} \cos (4 \beta ) +
       32 A_k k^2
        {{\lambda }^3} {{\mu }^4} \cos (4 \beta ) +
       128 A_{\lambda} {{\ {k}}^2} {{\lambda }^3}
        {{\mu }^4} \cos (4 \beta ) -
       2 A_{\lambda} {v^4} {{\lambda }^9} \cos (8 \beta ) + \nonumber \\
      & & 3 A_{\lambda} {v^4} {{\lambda }^7} {{\bar{g}}^2} +
      32 A_{\lambda} {v^2} {{\lambda }^5} {{\mu }^2}
        {{\bar{g}}^2} -
       16 A_k A_{\lambda} k
        {{\lambda }^2} {{\mu }^3} {{\bar{g}}^2} +
       64 k v^2 {{\lambda }^4}
        {{\mu }^3} {{\bar{g}}^2} -
       16 A_k k^2 \lambda
        {{\mu }^4} {{\bar{g}}^2} + \nonumber \\
       & & 64 A_{\lambda} k^2 \lambda  {{\mu }^4}
        {{\bar{g}}^2} + 64 k^3 {{\mu }^5}
        {{\bar{g}}^2} -
       4 A_{\lambda} {v^4} {{\lambda }^7} \cos (4 \beta )
        {\bar{g}^2} -
       32 A_{\lambda} {v^2} {{\lambda }^5} {{\mu }^2}
       \cos (4 \beta ) {\bar{g}^2} - \nonumber \\
      & & 16 A_k A_{\lambda} k
        {{\lambda }^2} {{\mu }^3} \cos (4 \beta )
        {\bar{g}^2} -
       64 k {v^2} {{\lambda }^4}
        {{\mu }^3} \cos (4 \beta )
        {\bar{g}^2} -
       16 A_k {k^2} \lambda
        {{\mu }^4} \cos (4 \beta )
        {\bar{g}^2} + \nonumber \\
      & & 64 A_{\lambda} {k^2} \lambda  {{\mu }^4}
        \cos (4 \beta ) {\bar{g}^2} +
       64 {k^3} {{\mu }^5}
        \cos (4 \beta ) {\bar{g}^2} +
       A_{\lambda} {v^4} {{\lambda }^7}
        \cos (8 \beta ) {\bar{g}^2} +
       48 {A_{\lambda}^2} {v^2} {{\lambda }^7} \mu
        \sin (2 \beta ) + \nonumber \\
      & & 24 A_k k {v^2}
        {{\lambda }^6} {{\mu }^2} \sin (2 \beta ) +
       120 A_{\lambda} k {v^2} {{\lambda }^6}
        {{\mu }^2} \sin (2 \beta ) +
       256 {A_{\lambda}^2} {{\lambda }^5} {{\mu }^3}
        \sin (2 \beta ) +
       96 {v^2} {{\lambda }^7} {{\mu }^3}
        \sin (2 \beta ) + \nonumber \\
     & &  768 A_{\lambda} k {{\lambda }^4} {{\mu }^4}
        \sin (2 \beta ) + 512 {k^2} {{\lambda }^3}
        {{\mu }^5} \sin (2 \beta ) -
       12 {A_{\lambda}^2} {v^2} {{\lambda }^5} \mu
        {{\bar{g}}^2} \sin (2 \beta ) - 12 A_k k {v^2}
        {{\lambda }^4} {{\mu }^2} {{\bar{g}}^2}
        \sin (2 \beta ) - \nonumber \\
      & & 60 A_{\lambda} k v^2 {{\lambda }^4}
        {{\mu }^2} {{\bar{g}}^2} \sin (2 \beta ) -
       16 k^2 v^2 {{\lambda }^3}
        {\mu }^3 {\bar{g}}^2 \sin (2 \beta ) -
       48 {v^2} {{\lambda }^5} {\mu }^3
        {{\bar{g}}^2} \sin (2 \beta ) -
       16 {A_{\lambda}^2} {v^2} {{\lambda }^7} \mu
        \sin (6 \beta ) - \nonumber \\
     & &  8 A_k k {v^2} {{\lambda }^6} {{\mu }^2} \sin (6 \beta ) -
       40 A_{\lambda} k {v^2} {{\lambda }^6} {{\mu }^2} \sin (6 \beta ) -
       32 {v^2} {{\lambda }^7} {{\mu }^3} \sin (6 \beta ) +
       4 {A_{\lambda}^2} {v^2} {{\lambda }^5} \mu
        {{\bar{g}}^2} \sin (6 \beta ) + \nonumber \\
    & &  \left. 4 A_k k {v^2}
        {{\lambda }^4} {{\mu }^2} {{\bar{g}}^2} \sin (6 \beta ) +
       4 A_{\lambda} k {v^2} {{\lambda }^4}
        {{\mu }^2} {{\bar{g}}^2} \sin (6 \beta ) -
       16 {k^2} {v^2} {{\lambda }^3}
        {{\mu }^3} {{\bar{g}}^2} \sin (6 \beta ) +
       16 {v^2} {{\lambda }^5} {{\mu }^3}
        {{\bar{g}}^2} \sin (6 \beta ) \right). \nonumber \\
\end{eqnarray}
All parameters were defined previously. Recall that
$\mu=\lambda x$.

\setcounter{equation}{0}
\section{Comments on Naturalness}
\label{commentonprob}

We have studied the NMSSM with extra vector-like quarks in
Section~\ref{extraquarks} and discussed that the model requires a delicate
cancellation among independent parameters.  In this appendix, we
make further comments on the naturalness of this model.

{}From Fig.~\ref{fig:2plots}, one can easily note that not only does the
experimentally allowed value of $v$ lie on a steep region of the
parameter space, which requires a degree of cancellation of order 1\%,
but it lies on the {\it steepest} region of the parameter space.

\begin{figure}[tb]
  \begin{center}
    \leavevmode
    \epsfxsize=0.6\textwidth \epsfbox{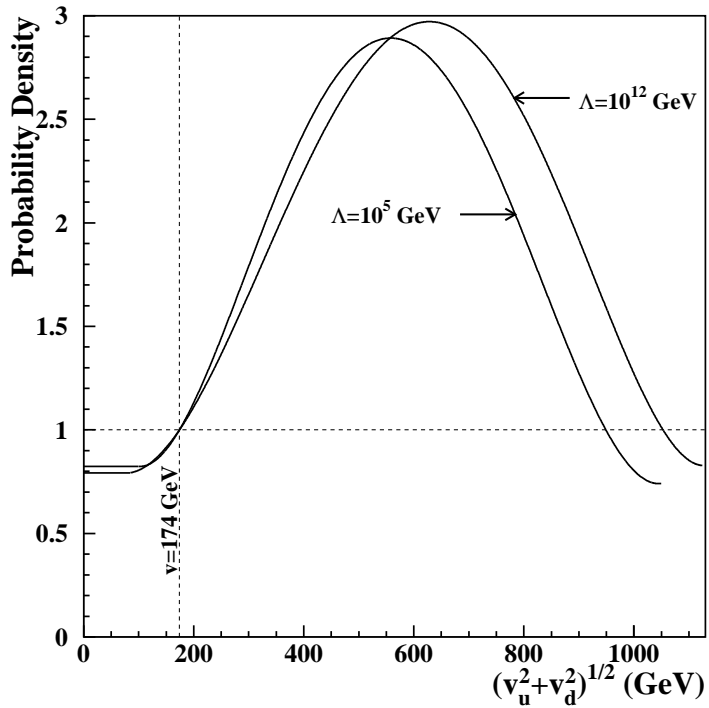}
    \caption[121]{The probability densities of finding specific values
      of $v$ in the NMSSM with extra vector-like quarks upon random
      choices of $\lambda$.  All other parameters are the same as in
      Fig.~\ref{fig:2plots}.  The probability densities are normalized
      so that $P(v=174~\rm{GeV})=1$.}
    \label{fig:prob}
  \end{center}
\end{figure}

One may, therefore, try to address the following question: if all
parameters are kept fixed (and this choice of parameters yields an
experimentally allowed spectrum) except one ({\it e.g.} $\lambda$),
what is the likelihood of obtaining a certain value of $v$ upon a random
choice of the 
free parameter? In other words, what is the probability $P(v)\,{\rm
  d}v$ of finding the
value of $\sqrt{v_u^2+v_d^2}$ between $v$ and $v+{\rm d}v$
given 
a random choice of $\lambda$?  This line of reasoning is
related to the definition of fine-tuning introduced by Anderson and
Casta\~no \cite{AC}.
It is easy to note that 
\begin{equation}
P(v)\propto \left({{\rm{d}} v\over \rm{d}\lambda}\right )^{-1}.
\end{equation}
This ``probability density'' is plotted in Fig.~\ref{fig:prob}. Note that we
restrict $\lambda$ to lie on a range where the same ``qualitative''
physics is obtained, that is, electroweak symmetry is broken and
$\tan(\beta)>1$. The plot has been normalized in such a way that
$P(v=174~{\rm GeV})=1$.

We note that, in some sense, the probability of living in our universe
is smaller, if this model is to be taken seriously, than the
probability of living in a universe where $v\simeq 600$ GeV by a factor of
three. One can turn this picture around and say that the NMSSM, with
the above choice of parameters, ``prefers'' (or predicts) $v\simeq 600$ GeV.

This does not happen in the MSSM. The analog of Fig.~\ref{fig:2plots} would be
Eq.~(\ref{eq:mueq}), which is a straight line ($M_{Z}^2=M_{Z}^2(\mu^2)$) 
if all parameters except
$\mu^2$ are kept fixed. In the language introduced above, the MSSM
does not ``prefer'' (or predict) any particular value of $M_{Z}^2$,
that is, the 
``probability density'' of $M_{Z}^2$ upon random choices of $\mu^2$ is flat.

\setcounter{equation}{0}
\section{The Dependence of the Higgs VEVs on the couplings of the
  modified NMSSM}
\label{estimatingsensitivity}

In Subsection \ref{extraquarks} we showed that the values of the Higgs
boson VEVs were extremely sensitive to small variations of the
superpotential couplings $\lambda$ and $k$. These variations were
evaluated numerically after Eq.~(\ref{variation}) for one particular set
of $\lambda$ and $k$. In this appendix we
study this issue analytically and show how one can estimate the
effects of small variations $\Delta\lambda$ and $\Delta k$ on $v$,
$\tan\beta$, and $x$. 

We will use the following three equations, derived in Section~\ref{sec:NMSSM}:
\begin{eqnarray}
 \label{d1} 
{\bar{g}^2 v^2 \over 4} &=& -\lambda^2 x^2+f\, , \\
 \label{d2}
2 k^2 x^2 &\simeq& -m_N^2+\lambda(k\sin2\beta-\lambda) v^2\, ,\\
\label{d3}
\sin2\beta &\simeq& 2 \frac{k}{\lambda}
\left[\frac{-m_{H_u}^2}{m_{H_d}^2-m_{H_u}^2}\right]\, ,
\end{eqnarray}
where
\begin{eqnarray}
f\equiv- \frac{1}{2} (m_{H_d}^2+m_{H_u}^2)
   -  \frac{1}{2} \frac {m_{H_d}^2-m_{H_u}^2}{\cos 2 \beta}=
 \frac{m_{H_d}^2-m_{H_u}^2 \tan^2 \beta}{\tan^2 \beta-1}\, ,
\end{eqnarray}
and we dropped the $A$-terms in Eq.~(\ref{d2}). Notice that, because
the term on the left-hand side of Eq.~(\ref{d1}) is much smaller then
each of the terms on the right-hand side, $f\simeq\lambda^2 x^2$.

For the purpose of the following estimates we will keep only the
largest terms in 
the variations. According to the numbers presented after
Eq.~(\ref{variation}), for a small variation of $\lambda$ the largest
variation on the right-hand side of Eq.~(\ref{d1}) is $2\lambda x^2
\Delta\lambda$. We can therefore write
\begin{equation}
\label{Dv}
\Delta (v^2) \approx -4{2\lambda x^2
  \Delta\lambda \over \bar{g}^2}\, .
\end{equation}
We will justify this approximation {\it a posteriori}.
Also, in our analysis we will completely neglect the dependence of the
soft-breaking masses-squared on $\lambda$ and $k$. This dependence is
very weak, as seen in the numbers presented after
Eq.~(\ref{variation}).

A small change in $\lambda$ results in a large change in
$v$. Hence, to determine the corresponding change in $x$, one can use
Eq.~(\ref{d2}) and only consider the variation of $v^2$, which is
approximately given by Eq.~(\ref{Dv}). We find
\begin{eqnarray}
4 k^2 x \Delta x \approx \lambda(k\sin2\beta-\lambda) \Delta
(v^2)\approx \lambda(k\sin2\beta-\lambda)(-4{2\lambda x^2
  \Delta\lambda \over \bar{g}^2})\, , 
\end{eqnarray}
so that
\begin{eqnarray}
{\Delta x \over x} \approx -\frac{2\lambda^3(k\sin2\beta-\lambda)}
{k^2\bar{g}^2} {\Delta\lambda \over \lambda}\, . 
\end{eqnarray}
For the point considered in the text 
($\lambda=0.11,\;k=-.045,\;\tan\beta=-2.9$) one finds 
$(\Delta x)/ x \approx 0.2 (\Delta\lambda)/ \lambda$.

Under a small change $\Delta k$, again using Eqs.~(\ref{d2},\ref{Dv}),
\begin{eqnarray}
4(k \Delta k x^2+ k^2 x \Delta x)\approx \lambda(k\sin2\beta-\lambda)
(-4{2\lambda^2 x \Delta x \over \bar{g}^2}).
\end{eqnarray}
Solving for $\Delta x/x$.
\begin{eqnarray}
{\Delta x \over x}  \approx {\Delta k \over k}
\left(-1-\frac{2\lambda^3(k\sin2\beta-\lambda)}{k^2\bar{g}^2}\right)^{-1}\, . 
\end{eqnarray}
Numerically, $(\Delta x)/ x \approx -1.2 (\Delta k)/ k$.

Next, we consider the effect of $\Delta\lambda$ on $f$. The problem comes
down to estimating $\Delta (\cos2\beta)^{-1}$, which can be done with
the aid of Eq.~(\ref{d3}): 
\begin{equation}
\Delta(\cos2\beta)^{-1}=-{\sin2\beta \over \cos^3 2\beta} \Delta (\sin
2\beta) \approx -{\sin2\beta \over \cos^3 2\beta}
\left(-2 {k\over\lambda^2} \frac{-m_{H_u}^2}{m_{H_d}^2-m_{H_u}^2}
\right) \Delta\lambda \, .
\end{equation} 
Thus,
\begin{equation}
 \label{d9}
\frac{\Delta f}{f}
\approx -{k\over\lambda} \frac{(-m_{H_u}^2)}{\lambda^2 x^2}
{\sin2\beta \over \cos^3 2\beta}{\Delta\lambda \over \lambda}\, .
\end{equation}
Plugging in the numerical values of the parameters, we find that the  
right-hand side of Eq.~(\ref{d9}) equals 
$-0.5 {\Delta\lambda / \lambda}$. Thus, a 1\% change in $\lambda$
results in a 0.5\% change in the value of $f$. Since $\lambda^2 x^2$
changes by 2\% in this case, the contribution of $f$ to the
variation of $v$ is approximately one fourth of that of $\lambda^2 x^2$,
consistent with the numbers given in Subsection \ref{extraquarks}.

The above argument can be repeated to find the effect of $\Delta k$ on
$f$. Notice that $\sin2\beta$ depends on the ratio $k/\lambda$
(Eq.~(\ref{d3})), and hence changing $k$ by $+1\%$ has the same effect
on $f$ as changing $\lambda$ by $-1\%$.

Finally, we show that the condition for $v$ to remain constant is 
$\Delta k/k=\Delta\lambda/\lambda$. We have already argued that $\sin
2\beta$, and therefore $f$, stays unchanged in this case and now show
that the same is true for $\lambda^2 x^2$.
Under $\lambda\rightarrow\lambda+\Delta\lambda $ the term $\lambda^2
 x^2$ changes by $2\lambda^2 x^2 ((\Delta\lambda/\lambda)+(\Delta x/x))=
2 \lambda^2 x^2 (1+0.2)(\Delta\lambda/\lambda)$,
 while under $k\rightarrow k+\Delta k $ it changes by  
$2 \lambda^2 x^2 (\Delta x/x)=
2\lambda^2 x^2 (-1.2)(\Delta k/k)$. These variations can be made to
cancel by imposing $\Delta k/k=\Delta\lambda/\lambda$.


\begin{thebibliography}{99}

\bibitem{SUSY1} S.~Dimopoulos and S.~Raby,\/ {\sl Nucl. Phys.} {\bf
    B192}, 353 (1981). 

\bibitem{SUSY2} E.~Witten, {\sl Nucl. Phys.}\/ {\bf B188}, 513 (1981).

\bibitem{SUSY3} M.~Dine, W.~Fischler, and M.~Srednicki, {\sl
    Nucl. Phys.}\/ {\bf B189}, 575
(1981).

\bibitem{radbreak} K.~Inoue, A.~Kakuto, H.~Komatsu and S.~Takeshita,
  {\sl Prog. Theor. Phys.}\/ {\bf 68}, 927 (1982);
  {\sl Prog. Theor. Phys.}\/ {\bf 71}, 413 (1984); \\
  L.E.~Ib\'a\~nez, {\sl Phys. Lett.}\/ {\bf B118}, 73 (1982); {\sl
    Nucl. Phys.}\/ {\bf B218}, 514 (1983); \\
  L.~Alvarez-Gaum\'e, J.~Polchinski and M.~Wise {\sl Nucl. Phys.}\/
  {\bf B221}, 495 (1983).

\bibitem{LEP2} R.~Akers {\it et al.}, CERN-PPE/97-083; \\
  P.~Abreu {\it et al.}, CERN-PPE/97-107; \\ 
  R. Barate {\it et al.}, CERN-PPE-97-128;\\ 
  M.~Acciari {\it et al.}, CERN-PPE/97-130.

\bibitem{DN} M. Dine and A.E. Nelson, {\sl Phys. Rev.}\/ {\bf D48}, 1277
  (1993), hep-ph/9303230.

\bibitem{DNS} M. Dine, A.E. Nelson and Y. Shirman, {\sl Phys. Rev.}\/
  {\bf D51}, 1362 (1995), hep-ph/9408384.

\bibitem{DNNS} M. Dine, A.E. Nelson, Y. Nir and Y. Shirman, {\sl Phys.
    Rev.}\/ {\bf D53}, 2658 (1996), hep-ph/9507378.
  
\bibitem{dynSUSY} 
  K.~Intriligator, N.~Seiberg and S.~Shenker, {\sl Phys. Lett.}\/ {\bf
    B342} 152 (1995), hep-ph/9410203; \\ 
  H. Murayama, {\sl Phys. Lett.}\/ {\bf B355}, 187 (1995), hep-th/9505082;\\
  E.~Poppitz and S. P.~Trivedi, {\sl Phys. Lett.}\/ {\bf B365}, 125
  (1996), hep-th/9507169;\\
  K.-I.~Izawa and T.~Yanagida, {\sl Prog. Theor. Phys.}\/
  {\bf  95}, 829 (1996), hep-th/9602180; \\
  K.~Intriligator and S.~Thomas, {\sl Nucl. Phys.}\/ {\bf B473}, 121
  (1996), hep-th/9603158; \\
  C.~Csaki, L.~Randall, and W.~Skiba, {\sl Nucl. Phys.}\/ {\bf B479}, 65
  (1996), hep-th/9605108; \\
  E.~Poppitz, Y.~Shadmi, and S.P.~Trivedi, {\sl Nucl. Phys.}\/ {\bf B480},
  125 (1996), hep-th/9605113; \\ 
  C.-L. Chou, {\sl Phys. Lett.}\/ {\bf B391}, 329 (1997),
  hep-th/9605119;\\
  T. Hotta, K.-I. Izawa, and T. Yanagida, {\sl Phys. Rev.}\/ {\bf D55},
  415 (1997), hep-ph/9606203;\\
  E.~Poppitz, Y.~Shadmi, and S. P.~Trivedi,
  {\sl Phys. Lett.}\/ {\bf  B388}, 561 (1996), hep-th/9606184; \\
  C.~Csaki, L.~Randall, W.~Skiba, and R.~G.~Leigh, {\sl Phys. Lett.}\/
  {\bf B387} 791 (1996), hep-th/9607021; \\ 
  K.~Intriligator and S.~Thomas, SLAC-PUB-7143, hep-th/9608046; \\
  E.~Poppitz and S.P.~Trivedi, {\sl Phys. Rev.}\/ {\bf D55},
  5508 (1997), hep-ph/9609529; \\
  C.~Csaki, M.~Schmaltz, and W.~Skiba, {\sl
    Phys. Rev.} {\bf D55}, 7840 (1997), hep-th/9612207; \\ 
  L. Randall, {\sl Nucl. Phys.}\ {\bf B495}, 37 (1997),
  hep-ph/9612426;\\
  N.~Haba, N.~Maru, and T.~Matsuoka, {\sl
    Nucl. Phys.}\/ {\bf B497}, 31 (1997), hep-ph/9612468; \\
  N. Arkani-Hamed, J. March-Russell, and H. Murayama,
  LBL-39865, hep-ph/9701286, to appear in {\sl Nucl. Phys.}\/ {\bf B}.
  E.~Poppitz, and S.P.~Trivedi, {\sl Phys. Lett.}\/ {\bf
    B401}, 38 (1997), hep-ph/9703246; \\
  N.~Haba, N.~Maru, and T.~Matsuoka, {\sl Phys. Rev.}\/ {\bf D56}, 4207
  (1997), hep-ph/9703250; \\ 
  Y.~Shadmi, {\sl Phys. Lett.}\/ {\bf B405}, 99 (1997), hep-ph/9703312; \\
  R. G. Leigh, L. Randall, and R. Rattazzi, {\sl Nucl. Phys.}, {\bf B501},
  375 (1997), hep-ph/9704246;\\
  K.-I. Izawa, Y. Nomura, K. Tobe, and T. Yanagida, {\sl Phys. Rev.}\/
  {\bf D56}, 2886 (1997), hep-ph/9705228;\\
  C. Csaki, L. Randall, and W. Skiba, MIT-CTP-2654, hep-ph/9707386;\\
  Y. Nomura, K. Tobe, and T. Yanagida, UT-797, hep-ph/9711220.

\bibitem{direct} H. Murayama, {\sl Phys. Rev. Lett.}\/ {\bf 79},
  18 (1997), hep-ph/9705271.

\bibitem{Amess} S. Dimopoulos, G. Dvali, R. Rattazzi, and G.F. Giudice,
  CERN-TH/97-98, hep-ph/9705307.

\bibitem{direct2}
  M. Luty, UMD-PP-97-132, hep-ph/9706554;\\ 
  M. Luty and J. Terning, MD-PP-98-98, hep-ph/9709306;\\ 
  Y. Shirman, PUPT-1721, hep-ph/9709383.



\bibitem{IS} K.~Intriligator and N.~Seiberg, {\sl
    Nucl. Phys. Proc. Suppl.}\/ {\bf 45BC}, 1 (1996), hep-th/9509066.
  
\bibitem{oldGM} C.~Nappi and B.~Ovrut, {\sl Phys. Lett.}\/ {\bf B113},
  175 (1982);\\
  M.~Dine and W.~Fischler, {\sl Nucl. Phys.}\/ {\bf B204}, 346 (1982);\\
  L.~Alvarez-Gaum\'e, M.~Claudson and M.~Wise, {\sl Nucl. Phys.} {\bf
    B207} 96 (1982).

\bibitem{Eetal} J. Ellis, J.F. Gunion, H. E. Haber, L. Roszkowski, and
  F. Zwirner, {\sl Phys. Rev.}\/ {\bf D39}, 844 (1989), and references
  therein.
  
\bibitem{hidden} R. Barbieri, S. Ferrara, and C. A. Savoy, {\sl
    Phys. Lett.}\/ {\bf 119B}, 343 (1982);\\
  A. H. Chamseddine, R. Arnowitt, and P. Nath, {\sl Phys. Rev.
    Lett.}\/ {\bf 49}, 970 (1982);\\
  L. J. Hall, J. Lykken and S.  Weiberg, {\sl Phys. Rev.}\/ {\bf D27},
  2359 (1983).

\bibitem{Polonyi} J. Polonyi, Budapest preprint, KFKI-93 (1977).

\bibitem{gaugino} P. Nilles, {\sl Phys. Lett.}\/ {\bf 115B}, 193
  (1982); {\sl Nucl. Phys.}\/ {\bf B217}, 366 (1983).

\bibitem{O'Rafeartaigh} L. O'Rafeartaigh, {\sl Nucl. Phys.}\/ {\bf
    B96}, 331 (1975).

\bibitem{SUGRA}
G.F. Giudice and A. Massiero, {\sl Phys. Lett.} {\bf B206} (1988) 480.

\bibitem{DG} S. Dimopoulous and H. Georgi, {\sl Nucl. Phys.}\/ {\bf
    B193}, 150 (1981).

\bibitem{NS} Y. Nir and N. Seiberg, {\sl Phys. Lett.}\/ {\bf B309},
  337 (1993), hep-ph/9304307.

\bibitem{non-abelian}   P. Pouliot and N. Seiberg, {\sl Phys.
    Lett.}\/ {\bf B318}, 169 (1993), hep-ph/9308363;\\
  D. B. Kaplan and M. Schmaltz, {\sl Phys. Rev.}\/
  {\bf D49}, 3741 (1994), hep-ph/9311281;\\
  L. J. Hall and H. Murayama, {\sl Phys. Rev. Lett.}\/, {\bf 75}, 3985
  (1995), hep-ph/9508296;\\
  C. D. Carone, L. J.  Hall, and H. Murayama, {\sl Phys.  Rev.}\/ {\bf
   D53}, 6282 (1996); {\it ibid}\/., {\bf D54}, 2328 (1996), hep-ph/9602364;\\
  R.  Barbieri, G. Dvali, and L.  J. Hall, {\sl Phys.  Lett.}\/ {\bf
    B377}, 76 (1996),  hep-ph/9512388;\\
  R. Barbieri, L.  J. Hall, S.  Raby, and A.  Romanino, {\sl Nucl.
    Phys.}\/ {\bf B493}, 3 (1997), hep-ph/9610449;\\
  R.  Barbieri, L. J. Hall, A.  Romanino, {\sl Phys. Lett.}\/ {\bf
    B401}, 47 (1997), hep-ph/9702315;\\
  P. H.  Frampton and O. C. W. Kong, {\sl Phys.  Rev.}\/ {\bf D55},
  5501 (1997); {\sl Phys. Rev. Lett.}\/ {\bf 77}, 1699 (1996), 
   hep-ph/9603372.

\bibitem{LNS} M. Leurer, Y. Nir, and N. Seiberg, {\sl Nucl. Phys.}\/
  {\bf B420} 468, (1994),  hep-ph/9310320.

\bibitem{Nir} Y. Nir, {\sl Phys. Lett.}\/ {\bf B354} 107, (1995), 
    hep-ph/9504312;\\ 
  Y. Nir and R. Rattazzi, {\sl Phys. Lett.}\/ {\bf B382},
  363 (1996), hep-ph/9603233.

\bibitem{KL} V. S. Kaplunovsky and J. Louis, {\sl Phys. Lett.}\/ {\bf
    B306}, 269 (1993), hep-th/9303040.

\bibitem{DDRandall} 
  N. Arkani-Hamed, C. D. Carone, L. J. Hall, and H. Murayama, {\sl
    Phys. Rev.}\/ {\bf D54}, 7032 (1996), hep-ph/9607298;\\
  I. Dasgupta, B. A. Dobrescu, and L. Randall, {\sl
    Nucl. Phys.}\/ {\bf B483}, 95 (1997), hep-ph/9607487.

\bibitem{DGP} G. Dvali, G.F. Giudice, and A. Pomarol, {\sl Nucl.
    Phys.}\/ {\bf B478} 31, (1996), hep-ph/9603238.

\bibitem{sliding} E.~Witten, {\sl Phys.  Lett.}\/ {\bf 105B}, 267
  (1981).

\bibitem{Nemeschansky} D. Nemeschansky, {\sl Nucl. Phys.}\/ {\bf
    B234}, 379 (1984).

\bibitem{CP} P. Ciafaloni and A. Pomarol, {\sl Phys. Lett.} {\bf
    B404}, 83 (1997), hep-ph/9702410.

\bibitem{AG}
  K. Agashe and M. Graesser, {\sl  Nucl. Phys.}\/ {\bf B507}, 3
  (1997), hep-ph/9704206.

\bibitem{GR}
  G.F. Giudice and R. Rattazzi, CERN-TH-97-145, hep-ph/9706540.

\bibitem{DDR} S. Dimopoulos, G. Dvali and R. Rattazzi, CERN-TH/97-186,
  hep-ph/9707537.

\bibitem{Y} T. Yanagida, {\sl Phys. Lett.} {\bf B400} 109, (1997), 
  hep-ph/9701394 .

\bibitem{NP} P. Nilles and N. Polonsky, TUM-HEP-278-97, hep-ph/9707249.

\bibitem{Sarid} R. Rattazzi and U. Sarid, {\sl Nucl. Phys.}\/ {\bf
    B501}, 297 (1997), hep-ph/9612464;\\ 
  F.M.  Borzumati, WIS-96/50/Dec.-PH, hep-ph/9702307.

\bibitem{BG} R. Barbieri and G.F. Giudice, {\sl Nucl. Phys.}\/ {\bf
    B306} 63 (1989).

\bibitem{Jerusalem} P. Janot, talk presented at {\sl E.P.S. HEP
    Conference}, Jerusalem (1997). \\ S. Assay, talk presented at {\sl
    E.P.S. HEP Conference}, Jerusalem (1997). \\ D. Buskulic {\it
    et al.}, {\sl Phys. Lett.}
  {\bf B373}, 246 (1996). 
  
\bibitem{D-term} K.R. Dienes, C. Kolda, and J. March-Russell, {\sl
    Nucl. Phys.}\/ {\bf B492}, 104 (1997), hep-ph/9610479. \\
  S. Dimopoulos, S. Thomas, and J.D. Wells, {\sl Nucl. Phys.}\/ {\bf
    B488}, 39 (1997), hep-ph/9609434.

\bibitem{NMSSMreg} T. Elliott, S.F. King, and P.L. White, {\sl
    Phys. Lett.}\/ {\bf B351}, 213 (1995), hep-ph/9406303;\\ 
   S.F. King and P.L. White, {\sl Phys. Rev.}\/ {\bf D52}, 4183
   (1995), hep-ph/9505326.

\bibitem{AC} G. Anderson and
  D. Casta\~{n}o, {\sl Phys. Lett.}\/ {\bf
    B347}, 300 (1995), hep-ph/9409419.

\end{thebibliography}
\end{document}